\documentclass[aps,pra,twocolumn,a4paper,notitlepage,superscriptaddress,floatfix,10pt]{revtex4-1}
\usepackage[pdftex]{graphicx,color}% Include figure files
\usepackage{dcolumn}% Align table columns on decimal point
\usepackage{bm}% bold math

\usepackage{amsfonts}
\usepackage{amssymb}
\usepackage{amsmath}
\usepackage{amsthm}

\usepackage{subfigure}
\usepackage{rotate}

\usepackage{pdfsync}

\usepackage[pdftex]{hyperref}

\newcommand{\noshow}[1]{}

\let\oldmarginpar\marginpar
\renewcommand\marginpar[1]{\-\oldmarginpar[\raggedleft\tiny #1]%
{\raggedright\tiny #1}}

\DeclareMathOperator{\Tr}{Tr}

\newcommand{\avg}[1]{\left\langle #1 \right\rangle}
\newcommand{\bra}[1]{\langle#1|}
\newcommand{\ket}[1]{|#1\rangle}

\newcommand{\be}{\begin{equation}}
\newcommand{\ee}{\end{equation}}
\newcommand{\ba}{\begin{eqnarray}}
\newcommand{\ea}{\end{eqnarray}}
\newcommand{\n}{\nonumber \\ }
\newcommand{\mac}{\mathcal}

\newcommand{\bit}{\begin{itemize}}
\newcommand{\eit}{\end{itemize}}

\newcommand{\goldchain}{Fibonacci chain~}
\newcommand{\gmean}{\varphi}

\begin{document}

\title{The eigenstate thermalization hypothesis in constrained Hilbert spaces: a case study in non-Abelian anyon chains}

\author{A. Chandran}
\affiliation{Perimeter Institute for Theoretical Physics, Waterloo, Ontario N2L 2Y5, Canada}   \email{achandran@perimeterinstitute.ca}
\author{Marc D. Schulz}
\affiliation{School of Physics and Astronomy, University of Minnesota, Minneapolis, Minnesota 55455, USA}
\author{F. J. Burnell}
\affiliation{School of Physics and Astronomy, University of Minnesota, Minneapolis, Minnesota 55455, USA}

\date{\today}

\begin{abstract}
Many phases of matter, including superconductors, fractional quantum Hall fluids and spin liquids, are described by gauge theories with constrained Hilbert spaces.
However, thermalization and the applicability of quantum statistical mechanics has primarily been studied in unconstrained Hilbert spaces.
In this article, we investigate whether constrained Hilbert spaces permit local thermalization.  Specifically, we explore whether the eigenstate thermalization hypothesis (ETH) holds in a pinned Fibonacci anyon chain, which serves as a representative case study.
We first establish that the constrained Hilbert space admits a notion of locality, by showing that the influence of a measurement decays exponentially in space.
This suggests that the constraints are no impediment to thermalization.
We then provide numerical evidence that ETH holds for the diagonal and off-diagonal matrix elements of various local observables in a generic disorder-free non-integrable model.
We also find that certain non-local observables obey ETH. 
\end{abstract}

\maketitle
\section{Introduction}

%Thermalization in isolated quantum systems
The development of synthetic quantum many-body systems has rejuvenated interest in the foundations of statistical mechanics.
In particular, when does an isolated quantum system locally equilibrate? 
Although the {\it global} unitary dynamics of such a system preserves all the information about the initial state, local subsystems can nevertheless forget their initial conditions and reach thermal equilibrium if the information about the initial state is spread over the entire system at long times \cite{Neumann:1929aa,Goldstein:2010qd}.
That is, every small subsystem sees the rest of the system as a thermal reservoir.
The general conditions for thermalization in quantum many-body systems is a long-standing question, see Refs.~\onlinecite{Gemmer:2004aa,Gomez:2011aa,Polkovnikov:2011ys,DAlessio:2015aa,Nandkishore:2015aa,Borgonovi:2016aa} for recent progress.

%Gauge theories everywhere
Thermalization has primarily been explored in systems with a local tensor product structure, like spin chains, bosonic and fermionic systems \footnote{Despite the anti-commuting algebra obeyed by fermionic creation and annihilation operators, the fermion occupation on a given site is, at the level of the Hilbert space, independent of that of all other sites.}, through numerical studies \cite{Jensen:1985aa,Rigol:2008bh,Rigol:2009bh,Rigol:2010aa,Biroli:2010kl,Rigol:2012aa,Genway:2012aa,Khatami:2012aa,Steinigeweg:2013aa,Beugeling:2014aa,Sorg:2014aa,Kim:2014aa,Beugeling:2015aa} and experiments in few-body systems \cite{Neill:2016aa,Kaufman:2016aa}.
However, many phases of matter, including superconductors, fractional quantum Hall fluids and spin liquids, are described at low energies by gauge theories with no local tensor product structure.
The gauge symmetry imposes local constraints which lead to equilibrium properties disallowed in unconstrained models, for example, first order phase transitions \cite{Laumann:2012ab} and topological order in 1D.
Can the constraints also affect the ability of the system to thermalize under its own dynamics?
In this article, we focus on this question using a pinned non-Abelian anyon chain of Fibonacci anyons as a case study.

%Non-abelian Anyons
Particles with non-Abelian statistics arise in two spatial dimensions and have attracted significant interest recently due to their potential for robustly storing and processing quantum information \cite{KitaevTQC,Freedman2002,Stern:2010aa}.
The adiabatic exchange of such non-Abelian anyons entails a non-Abelian (i.e. matrix-valued) unitary transformation on the global state of the system, in contrast to the signs accumulated by conventional fermions and bosons.
In addition to being realizable in 2D topological superconductors and certain fractional Hall \cite{MooreRead,ReadRezayi} and spin liquid \cite{KitaevExact} states, there are a number of promising proposals \cite{KitaevMajorana,FuKaneMajorana,OregPRL105.177002,LutchynPRL105.077001,AliceaNatPhys7,ShibaStates,BarkeshliQi,LindnerParafermion,KirillParafermion,ChengParafermion} to engineer non-abelian anyons in quasi-1 dimensional systems. 
The non-Abelian statistics are typically encoded in a non-Abelian gauge theory; for example, the Fibonacci anyons that we study in this article are excitations of an appropriate Chern-Simons gauge theory.
Our main interest in this model stems from its local constraints: the Hilbert space of the Fibonacci chain is obtained from an Ising chain by projecting out specific Ising patterns locally.  

%Expectations about non-Abelian anyons
Previous studies of pinned non-Abelian anyons models, particularly in the strongly disordered context \cite{Fidkowski:2009aa,Kraus:2011aa,Laumann:2012aa,Laumann:2012ac,Vasseur:2015ab,Potter:2016aa}, suggest that it is hard to localize energy in anyon chains.
Indeed, Ref.~\onlinecite{Laumann:2012ac} argued that disordered Majoranas in 2D form a thermal metal, while one of the results of Ref.~\onlinecite{Potter:2016aa} is that the Fibonacci chain cannot be localized at any temperature.
The difficulty in localizing energy even in strongly disordered anyon chains suggests that these systems thermalize under their own dynamics.

In this article, we substantiate this intuition by showing that a conjectured description of thermalization, the eigenstate thermalization hypothesis (ETH), holds in a clean chain of pinned non-Abelian anyons with a non-integrable Hamiltonian.  
The ETH \cite{Jensen:1985aa,Deutsch:1991ss,Srednicki:1994dw,Srednicki:1999bd,Rigol:2008bh} asserts that thermalization occurs at the level of individual eigenstates.
Thus, eigenstate expectation values of few-body observables are given by expectation values in the thermal/Gibbs ensemble.
ETH also states that the off-diagonal matrix of few-body operators in the energy eigenbasis is a random matrix in sufficiently small energy windows \footnote{The energy window can at most polynomially vanish with system size}.
Below, we verify both properties.

Specifically, we provide two pieces of evidence in favor of thermalization.
First, we show that the influence of local measurements is quasi-local (i.e. exponentially localized) in the constrained Hilbert space.
This is reminiscent of the situation in 2D quantum Hall systems: strictly local operators projected into the lowest Landau level are Gaussian localized.  
As the constraints do not destroy spatial locality, we expect ETH to hold in the anyon chains. 
Note that thermalization in an Ising chain in which local Ising patterns are energetically penalized (as opposed to projected out) does not guarantee ETH in the projected subspace, as at any finite temperature, there is a finite density of the violating patterns.

%Numerical studies
Next, we provide numerical evidence obtained by exact diagonalization that a non-integrable Fibonacci anyon chain satisfies both diagonal and off-diagonal ETH at infinite temperature.
We focus on two kinds of observables: local observables that involve two or three nearby anyons and non-local observables (which are nonetheless quasi-local in the sense described above) that involve half the anyons in the chain. 
Although ETH is expected only to apply to the former, we will show that both kinds of observables thermalize, though the time-scale for the thermalization of the non-local observable is longer than the local ones.
In addition, we study diagonal ETH for a truly non-local observable, involving braiding an anyon around a fraction of the total anyons in the system.
Even here, we find that the behavior of this observable in an ensemble of eigenstates is captured by an ensemble of random vectors in the Hilbert space, suggesting that no vestige of the long-ranged entanglement used in topological quantum computing is present at high temperature in these systems.

The plan of the paper is as follows. In Sec.~\ref{Sec:Background}, we review the basics of Fibonacci anyons and define ETH for this system. 
In Sec.~\ref{CountingSec}, we show that the effect of a local measurement in the Fibonacci anyon Hilbert space decays exponentially away from the measurement point and derive thermal expectation values of various observables in these chains. 
We then turn to the numerical results in Sec.~\ref{Sec:Numerics} to argue in favor of ETH. 
We end with a discussion of thermalization in other non-Abelian anyon chains and in higher dimensions.

\section{Background}
\label{Sec:Background}

We review the two important ingredients of our study: 1D Fibonacci anyon chains, and the Eigenstate Thermalization Hypothesis.  
Readers familiar with either may wish to briefly skim this section for our notation, and proceed directly to Sec.~\ref{Sec:ThermalizationFib}.

\subsection{Fibonacci anyons}
\label{Sec:FibChainIntro}
Our study will focus on a particular 1D model of pinned interacting non-abelian anyons which we call the Fibonacci chain.  
This is an extension of the Golden chain model introduced in Ref.~\onlinecite{Feiguin:2007aa} with dimerized two-body and three-body interactions.
As we discuss in Sec.~\ref{Sec:DiscussionSection}, we expect that thermalization in this model  -- and the potential barriers thereto -- are representative of the behavior of more complex Hilbert spaces with local constraints.  
In addition, this model has the practical advantage of being one dimensional with a small effective local Hilbert space dimension ($\approx 1.6$), making it amenable to numerical study.

\begin{figure}[h!]
\begin{minipage}{\columnwidth}
$\begin{array}{c} \text{(a)}\\ \vspace{3em} \end{array}$
\includegraphics[width=0.5\textwidth]{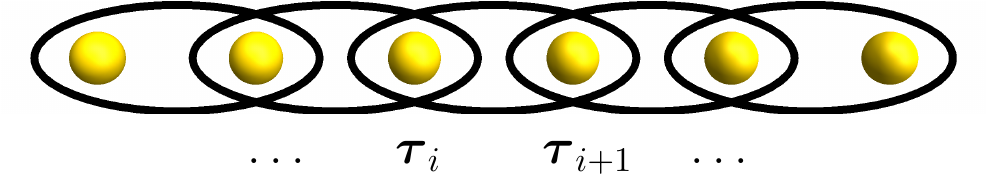}
\end{minipage}
\begin{minipage}{\columnwidth}
$\begin{array}{c} \text{(b)}\\ \vspace{5em} \end{array}$\includegraphics[width=0.75\textwidth]{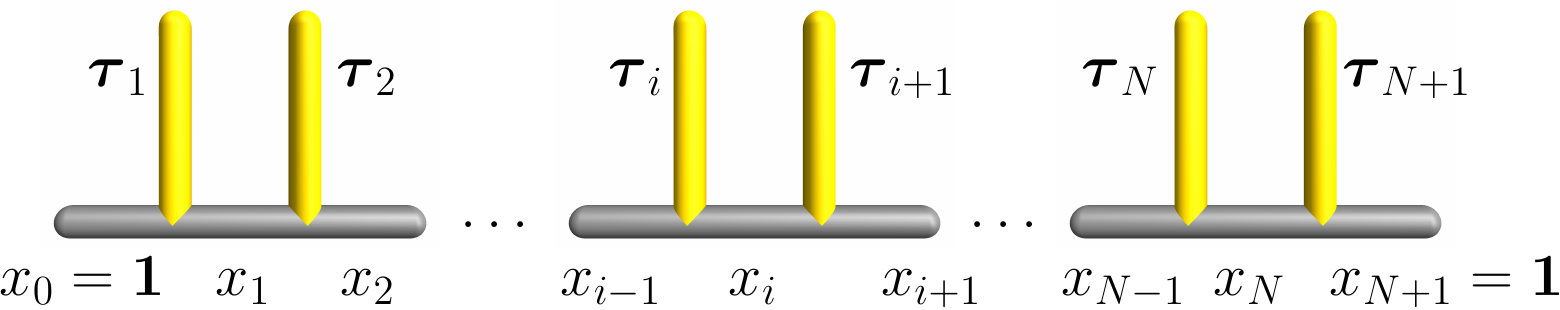}\vspace{-2em}
\end{minipage}
\caption{\label{GCFig} (a) A series of $N+1$ quasiparticles in a 2D anyon system arranged in a line.  Local measurements include probing the fusion channel of adjacent pairs of anyons, indicated here by black circles.  (b) The Hilbert space for the Fibonacci chain with $N$ bonds.  Vertical legs represent the $N+1$ anyons depicted in (a), while each horizontal bond represents the net fusion channel of all anyons to its left.}
\end{figure}

The fundamental degrees of freedom of the \goldchain are {\it Fibonacci anyons}, which can 
arise as low-energy quasiparticles in strongly correlated 2D systems \cite{slingerland01}.  
Fibonacci anyons have two defining properties.
First, any pair of Fibonacci anyons has a net anyonic charge, which can be either $1$ (meaning that the two anyons can be annihilated, leaving no particles behind) or $\tau$ (meaning that if the two anyons are brought close together, they will form a single anyon of the same type).  
This total anyonic charge  -- known as the {\it fusion channel} of the pair -- is reminiscent of the total spin of a pair of particles; we represent the possibilities compactly through the following {\it fusion rules} \cite{rowell09}:
\be \label{FuseEq}
\tau \times \tau = 1 + \tau \ , \ \ \ \tau \times 1 = \tau \ , \ \ \  1 \times 1 = 1 \ \ .
\ee
Unlike spin, however, the total anyonic charge of any number of Fibonacci anyons necessarily takes on one of only two values, $1$ or $\tau$.  
The second property is that exchanging a pair of the $\tau$ anyons results in a net phase that depends on their fusion channel:
\be
\label{Eq:Rmatdef}
R^{\tau\tau}_1 = e^{ 4 \pi i /5}  \ , \ \ \ R^{\tau\tau}_\tau = e^{ -3 \pi i /5}.  
\ee
where the subscript denotes the fusion channel of the two $\tau$ anyons.
Together, these two properties make Fibonacci anyons qualitatively different from particles with spin.  
Under exchange they are neither fermions nor bosons, but rather non-abelian anyons.  

The \goldchain can arise in a two dimensional system with anyonic excitations if the anyons are arranged in a line (Fig.~\ref{GCFig}(a)).  
This would be natural, for example, if the anyons are bound states \cite{FuKaneMajorana,KirillParafermion,LindnerParafermion} at the edges of a 2D topologically ordered system. 
In Fig.~\ref{GCFig}(b), the vertical legs represent the anyons shown in (a); the anyonic charge of each leg is always $\tau$.  
The $i^{th}$ horizontal bond encodes the net fusion channel of the first $i$ anyons in the chain, which can take on the values $X_i=1$ or $\tau$.  
The $X_i$ are thus Ising variables.
The Hilbert space of the \goldchain consist of all possible assignments of $X_i$ consistent with the fusion rules 
(\ref{FuseEq}).  
Since the trivial anyonic charge combined with the $\tau$ anyon always gives a $\tau$ anyon, the Hilbert space consists of all assignments of $X_i$ obeying the constraint that no two consecutive bonds are labelled by $X_{i}= 1$.  
The Hilbert space of the Fibonacci chain can thus be obtained from an Ising chain by projecting out all Ising configurations with $X_i = X_{i+1}=1$ for any $i$. 
With the open boundary conditions shown in Fig.~\ref{GCFig}, the number of allowed configurations of an $N$-bond chain grows as
\be \label{HSize}
n_N = f_{N}\stackrel{N \rightarrow \infty}{ \approx} \gmean^{N-1}
\ee
where $f_k$ is the $k^{th}$ Fibonacci number, and 
\be
\gmean = \frac{1}{2} \left( 1 + \sqrt{5} \right )
\ee
is known as the {\it Golden mean}.
Note that $X_0=X_{N+1}=1$ as the chain of anyons is assumed to be indistinguishable from the vacuum far away from the chain.  
Throughout this work, $N$ denotes the number of bonds in the chain excluding the boundaries; hence a system with $N$ bonds, {\it excluding} the initial and final bonds $X_0$ and $X_{N+1}$, consists of $N+1$ anyons.

\begin{figure}[h!]
\begin{tabular}{lll}
$\begin{array}{c} \text{(a)}\\ \vspace{3em} \end{array}$ $\begin{array}{c}\includegraphics[width=0.13\textwidth]{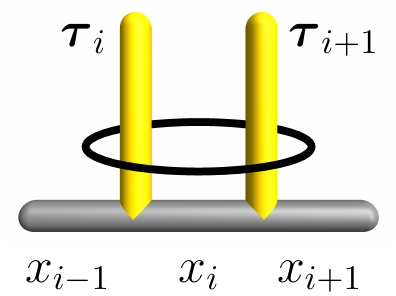}\end{array}$ & $= \sum\limits_{Y_i = 1, \tau} \alpha(Y_i)\hspace{-.75em} $&$\begin{array}{c}\includegraphics[width=0.13\textwidth]{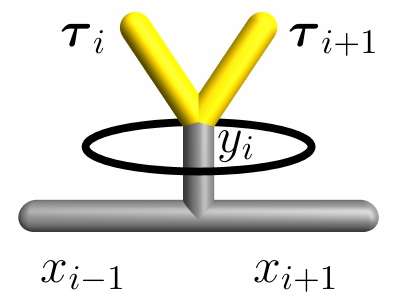}\end{array}$ \\ \vspace{.25em} & & \\
$\begin{array}{c} \text{(b)}\\ \vspace{3em} \end{array}$ $\begin{array}{c}\includegraphics[width=0.13\textwidth]{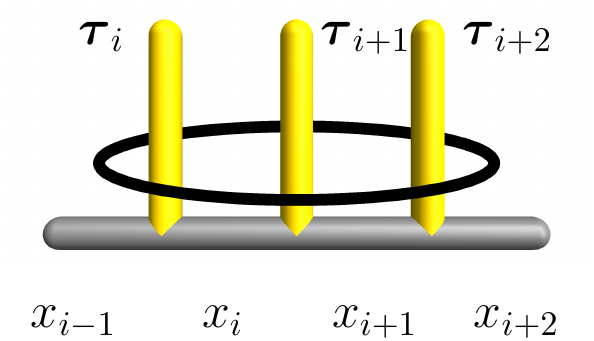}\end{array}$ & $= \sum\limits_{Y_i ,Z_i} \alpha(Y_i , Z_i) \hspace{-.75em}$ &$\begin{array}{c}\includegraphics[width=0.13\textwidth]{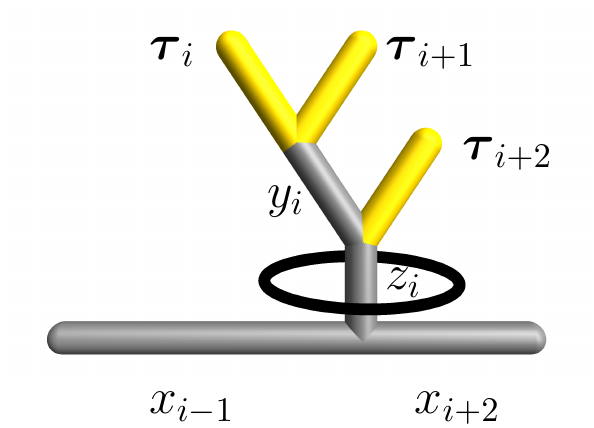}\end{array}$ \\
\end{tabular}
\caption{\label{HFig1} (a) A natural choice of Hamiltonian consists of energetically favouring one of the two fusion channels $Y_i = 1, \tau$ for each pair of neighbouring anyons.  (b) Another possible interaction measures the net fusion channel $Z_i = 1, \tau$ of each triple of neighbouring anyons. In both cases the dependence of the coefficients $\alpha$ on the labels $\lbrace X_j \rbrace$ is left implicit.}
\end{figure}

In the Golden chain model \cite{Feiguin:2007aa}, the interaction between the Fibonacci anyons have the same form as the Heisenberg interaction between spin-1/2 particles.
The Heisenberg interaction assigns an energy depending on the overall state of the two spins; similarly, the interactions in the Golden chain assigns an energy dependent on the fusion channel of the two anyons.
Assuming that the interactions between nearest neighbour anyons dominate, the Hamiltonian is:
 \be \label{HGC0}
H_0 = -  J \sum_{i=1}^{N} \Pi^1_{i,i+1}
\ee
where $\Pi_{i,i+1}^1$ is the projector onto configurations where the fusion channel of the pair of anyons at $i$ and $i+1$ is one.  
Let $Y_i$ denote the fusion channel of the anyons $i$ and $i+1$. 
$Y_i$  can be represented as a matrix mapping the five triples $(X_{i-1}, X_{i}, X_{i+1} ) = (1, \tau, 1), (\tau, \tau, 1), (1, \tau, \tau), (\tau, 1, \tau)$, and $(\tau, \tau, \tau)$ allowed by the constraint onto the five possible states 
$(X_{i-1}, Y_{i}, X_{i+1} ) = (1, 1, 1), (\tau, \tau, 1), (1, \tau, \tau), (\tau, 1, \tau)$, and $(\tau, \tau, \tau)$: 
\be\label{Ytrans}
Y_i = 
\begin{pmatrix}
1& 0&0 &0 &0 \\
0& 1& 0&0 &0 \\
0& 0 & 1& 0&0 \\
0 &0 &0 &\gmean^{-1}  &\gmean^{-1/2} \\
0 &0 &0 &\gmean^{-1/2} & -\gmean ^{-1} \\
\end{pmatrix},
\ee
see Fig.~\ref{HFig1}(a). 
The projector $\Pi_{i,i+1}^1$ is constructed by first making this change of basis, projecting onto those configurations with $Y_i =1$, and then inverting the basis transformation.  Expressed in terms of its action on the 5 triples $(X_{i-1}, X_{i}, X_{i+1} ) $, this gives:
\be \label{YProj}
\Pi_{i;2}^1\equiv \Pi_{i,i+1}^1 = 
 \begin{pmatrix}
1& 0&0 &0 &0 \\
0& 0& 0&0 &0 \\
0& 0 & 0& 0&0 \\
0 &0 &0 &\gmean^{-2}  &\gmean^{-3/2} \\
0 &0 &0 &\gmean^{-3/2} & \gmean^{-1} \\
\end{pmatrix}.
\ee
For notational brevity, we denote $\Pi_{i,i+1}^1$ by $\Pi_{i;2}^1$ henceforth. The subscript $2$ indicates that the projector acts on two neighbouring anyons.

However, the Hamiltonian (\ref{HGC0}) is integrable \cite{trebst08}.
Consequently, the Golden chain has an extensive number of conservation laws in addition to the total energy. 
As we are interested in the thermalization behavior of generic anyon chains, we deform $H_0$ in two ways to break integrability and other spatial symmetries.
First, we include a 3-body term $\Pi^1_{i; 3} \equiv \Pi^1_{i,i+1,i+2}$, which projects the net fusion channel $Z_i$ of three consecutive anyons to the trivial channel \cite{trebst08}. 
%We denote $\Pi^1_{i,i+1,i+2}$ by $\Pi^1_{i; 3}$ henceforth.
This operator can be expressed in terms of quadruples $(X_{i-1}, X_{i}, X_{i+1}, X_{i+2} )$ by making two subsequent basis transformations of the form Eq.~\eqref{Ytrans}, projecting onto states with $Z_i=1$, and then inverting the basis transformation (see Appendix \ref{3BodyApp} for details).  
The process is shown in Fig.~\ref{HFig1}(b).
Second, we dimerize the chain so that the couplings on bond $i$ depend on the parity of $i$.  
The net Hamiltonian is thus:
\begin{align} \label{HChain}
H &=  \sum_{i=1}^N  (\cos \theta_{e} \Pi^1_{2i;2} + \cos \theta_{o} \Pi^1_{2i-1;2} ) \nonumber \\
&+ \sum_{i=1}^{N-1} (\sin \theta_e \Pi^1_{2i;3} + \sin \theta_o  \Pi^1_{2i-1;3})
\end{align}
where we have set the overall scale of the Hamiltonian to be one.
Ref.~\onlinecite{Kakashvili:2012aa} describes the phase diagram when $\theta_e=\theta_o$ in detail.
The special integrable points in the phase diagram are at $\theta_e=\theta_o = 0, \pi$ and at $\tan(\theta_e)=\tan(\theta_o)=1/\varphi$.
Away from these points, $H$ is non-integrable, but possesses the extra symmetry of inversion.
It is simplest to break all spatial symmetries of the system to discuss thermalization; in the main text, we therefore present data at $\theta_e \neq \theta_o$ for open chains.
In Appendix~\ref{App:InversionSymm}, we present data in the inversion-symmetric open chain at $\theta_e=\theta_o$ for completeness.

We note that periodic boundary conditions allow access to larger system sizes as the numerical diagonalization can be performed within each momentum sector.
However, the spectrum within each momentum sector must be further resolved by its eigenvalue under a non-local string operator that commutes with the Hamiltonian, dubbed a ``topological symmetry" by Ref.~\onlinecite{Feiguin:2007aa}.
Since constructing this non-local string operator significantly increases the coding complexity, we leave this study for future work.

\begin{figure}[h!]
%$\begin{array}{c}  \text{(a)}\\ \vspace{5em} \end{array}$
\includegraphics[width=0.33\textwidth]{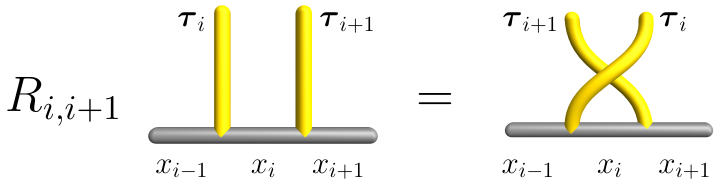}\\ \vspace{-1em}
\caption{\label{BraidFig} 
 The local operation $R$ which exchanges a pair of neighbouring anyons. 
 }
\end{figure}

In the following, we examine the eigenstate expectation values of local observables like $\Pi^1_{i;2}$ and $\Pi^1_{i;3}$ and non-local observables like the projector $\Pi^1_{i;x}$ onto $X_i=1$.
We also consider a third class of observables which cannot be constructed locally from a change of basis in either the $X$ or $Y$ variables.  
One example is the operator which braids an anyon around $l$ consecutive anyons. 
To construct this operator, we begin with an operator $R_{i, i+1}$ that exchanges the anyons on sites $i$ and $(i+1)$, as shown in Fig.~\ref{BraidFig}. %(a).  
In the basis $(X_{i-1}, X_{i}, X_{i+1} ) = (1, \tau, 1), (\tau, \tau, 1), (1, \tau, \tau), (\tau, 1, \tau)$, and $(\tau, \tau, \tau)$, it is given by the matrix \cite{Trebst:2008qf} 
\be
R_{i,i+1} = \begin{pmatrix}
e^{ 4 \pi i/5} & 0 & 0 & 0 & 0 \\
0 & e^{ -3 \pi i/5} & 0 & 0 & 0  \\
0 &0 & e^{ -3 \pi i/5} & 0 &  0  \\
0 & 0 & 0 &\frac{e^{ - 4 \pi i/5}}{\gmean}  & \frac{-  e^{ -2 \pi i/5} }{\sqrt{\gmean} }\\ 
0 & 0 & 0 &\frac{-  e^{ -2 \pi i/5} }{\sqrt{\gmean} }  & \frac{-1}{\gmean} \\
\end{pmatrix} .
\ee
Performing subsequent exchanges on a series of $l$ sites moves an anyon from site $i$ to site $i+l$.
The braid of the anyon $i$ around $l$ consecutive anyons is consequently performed by the product operator:
\begin{align} \label{LongBraid}
B_{i, i+l} &=& R_{i,i+1} R_{i+1, i+2}, ... R_{i+l-2, i+l-1}\\ \nonumber
&& \times R^2_{i+l-1, i+l}   ... R_{i+1, i+2} R_{i,i+1} \ \ .
\end{align}
The braid operator $B_{i,i+l}$ is a $f_{l+4} \times f_{l+4}$ matrix in the basis $(X_{i-1}, \ldots, X_{i+l} )$ and is non-local as it involves operators on all the bonds between $i-1$ and $i+l$.

\subsection{Eigenstate thermalization hypothesis}
\label{Sec:ETHIntro}
The foundations of statistical mechanics are built on the multitude of microstates that correspond to a single macroscopic state.
As the microstates are indistinguishable by local or few body measurements, the simplest assumption is that the stationary state of the system is an equiprobable mixture of all accessible microstates.
The eigenstate thermalization hypothesis (ETH) is the same hypothesis for quantum systems \cite{Jensen:1985aa,Deutsch:1991ss,Srednicki:1994dw,Srednicki:1999bd,Rigol:2008bh}.
It states that the stationary states of a quantum system, i.e. the many-body eigenstates, are locally indistinguishable from one another and are equal weight superpositions of all accessible microstates.
ETH was posited from information theoretic and semi-classical arguments nearly two decades ago \cite{Jensen:1985aa,Deutsch:1991ss,Srednicki:1994dw,Srednicki:1999bd}, and has been numerically tested in one and two dimensions \cite{Rigol:2008bh,Rigol:2009bh,Rigol:2010aa,Biroli:2010kl,Rigol:2012aa,Genway:2012aa,Khatami:2012aa,Steinigeweg:2013aa,Beugeling:2014aa,Sorg:2014aa,Kim:2014aa,Beugeling:2015aa}.

Although initially proposed for systems with well-defined local Hilbert spaces, the ETH ansatz can be easily adapted to the non-Abelian anyonic systems of interest in this article.
Concretely, consider a non-integrable Hamiltonian $H$ of an anyonic system with eigenstates $\ket{n}$ at energies $E_n$.
Let $\mathcal{O}$ denote a local or a few-body operator, for example, $\Pi^1_{i;2}$ or $B_{i,i+1}$.
ETH prescribes the diagonal and off-diagonal structure of the matrix elements of $\mathcal{O}$ in the energy eigenbasis.
Mathematically:
\begin{align}
\label{Eq:ETHansatz}
\bra{m}\mathcal{O}\ket{n} = \mathcal{O}(\bar{E}) \delta_{mn} + \frac{r_{mn}}{\sqrt{e^{S(\bar{E})} }} f_{\mathcal{O}}(\bar{E}, E_m - E_n)
\end{align}
where $ \mathcal{O}$ and $f_{\mathcal{O}}$ are smooth functions of their arguments on the scale of the many-body level spacing, $\delta_{mn}$ is the Kronecker delta function, $r_{mn}$ are independent Gaussian distributed random variables with zero mean and unit variance, $\bar{E} = (E_m + E_n)/2$ is the average energy and $S(E)$ is the thermal entropy at energy $E$.

Every term in Eq.~\eqref{Eq:ETHansatz} needs explanation.
Recall that the expectation value of $\mathcal{O}$ in the canonical ensemble is:
\begin{align}
\label{Eq:CanEnsembleOExp}
\avg{\mathcal{O}}_c \equiv \frac{\Tr e^{-\beta H} \mathcal{O}}{\Tr e^{-\beta H}},
\end{align}
Following Ref.~\onlinecite{Srednicki:1999bd}, we subsitute Eq.~\eqref{Eq:ETHansatz} in the above expression and obtain:
\begin{align}
\label{Eq:OOcanFiniteN}
\mathcal{O}(\bar{E}) = \avg{\mathcal{O}}_c + O(N^{-1}),
\end{align}
where $N$ is the number of particles and $\bar{E}$ is the average energy at inverse temperature $\beta$: $\bar{E} = \Tr H e^{-\beta H}/\Tr e^{-\beta H}$.
Note that we have ignored the correction due to the second term in Eq.~\eqref{Eq:ETHansatz}, which decreases exponentially with $N$.
In the thermodynamic limit:
\begin{align}
\mathcal{O}(\bar{E}) = \avg{\mathcal{O}}_c, \quad N \to \infty,
\end{align}
so that $\bra{n} \mathcal{O} \ket{n}$ is independent of the eigenstate index $n$ for all eigenstates with the energy density $\bar{E}/N$ and equals the expectation value in the canonical ensemble.
By the equivalence of ensembles, $\mathcal{O}(\bar{E})$ also equals the expectation value in the microcanonical ensemble in the thermodynamic limit.

At finite system size, the difference between $\mathcal{O}(\bar{E})$ and $ \avg{\mathcal{O}}_c$ is $O(N^{-1})$ (Eq.~\eqref{Eq:OOcanFiniteN}).
However, Eq.~\eqref{Eq:ETHansatz} posits that the difference between expectation values taken in eigenstates with nearby energies is much smaller, and is a Gaussian random variable with variance proportional to $e^{-S(E_n)}$.
As $S(E_n)$ is proportional to the number of anyons $N$ at any energy density $E_n/N$, the distribution of the differences vanishes exponentially with $N$.
There is numerical evidence that the extreme value statistics of the distribution also obeys Eq.~\eqref{Eq:ETHansatz}, see Ref.~\cite{Kim:2014aa} for more details.

The off-diagonal structure of the matrix elements is encoded by the second term in Eq.~\eqref{Eq:ETHansatz}.
At finite size $N$, the smooth function $f_{\mathcal{O}}(\bar{E}, \omega)$ may be replaced by a constant in a sufficiently small energy window about $E_n$ (we discuss why in Sec.~\ref{Sec4B}).   
Then, the second term in Eq.~\eqref{Eq:ETHansatz} implies that $\mathcal{O}\ket{n}$ is as random as can be subject to the energy constraint.
That is, $\mathcal{O}\ket{n}$ is a random vector in this energy window.
This also explains the factor $1/\sqrt{e^{S(\bar{E})} }$: it is required for the normalization of the random vector.
The function $f_{\mathcal{O}}(\bar{E}, \omega)$ is related to the spectral density of the operator $\mathcal{O}$:
\begin{align}
\label{Eq:fSpecDensity}
\int_{-\infty}^{\infty} dte^{-i\omega t} \bra{n} \mathcal{O}(t) \mathcal{O}(0) \ket{n}_c = 2\pi |f_{\mathcal{O}}(\bar{E}, \omega)|^2 e^{-\beta \omega/2}
\end{align}
where the correlator in the LHS is connected and $\omega$ is assumed to be much smaller than $E_n$, so that $\bar{E} \approx E_n$ (see Ref.~\cite{DAlessio:2015aa} for details).
In thermalizing systems, the spectral density and $f_\mathcal{O}$ are well-defined in the thermodynamic limit, with the Kubo formula relating the spectral density to linear response.
We can therefore infer general properties of $f_{\mathcal{O}}$ from the properties of the associated linear response susceptibility.
We discuss and test some of these properties in Sec.~\ref{Sec4B}.
%As the Kubo formulae determine linear response, it follows from Eq.~\eqref{Eq:fSpecDensity} that individual eigenstates and the canonical ensemble have the same linear response.

It is the goal of this article to check Eq.~\eqref{Eq:ETHansatz} for Fibonacci anyon chains.
A number of properties of the anyonic chain follow from Eq.~\eqref{Eq:ETHansatz}.
First, local observables reach their thermal values at late times starting from any initial state $\ket{\psi}$.
Thus, the anyonic chain acts as its own bath in isolation.
Next, the time-scales for local relaxation can be extracted from the Fourier transform of $f_{\mathcal{O}}$ function (Eq.~\eqref{Eq:fSpecDensity}).

We note that ETH is posited only for local observables like $\Pi^1_{i;2}$.
Thus, only the expectation values of local observables approach thermal values at late times irrespective of the starting state $\ket{\psi}$.
Non-local observables like $\Pi^1_{N/2; x}$ or $B_{1,N/2}$ that act on all $N/2$ anyons in the left half of the chain need not thermalize.
In particular, their expectation values can differ between adjacent eigenstates in the spectrum.
In the numerical study, we will however find that this is not the case.

 \section{Properties of the thermal ensemble in the \goldchain }
\label{Sec:ThermalizationFib}
As described in Sec.~\ref{Sec:FibChainIntro}, the \goldchain Hilbert space can be obtained from that of the Ising model by projecting out configurations in which two consecutive bond variables obey $X_i = X_{i+1} = 1$.  Consequently, unlike the Ising model, 
the anyonic model in Eq.~\eqref{HChain} does not admit a simple tensor product Hilbert space.

In this section, we show that the lack of a local product structure means that the measurement of observables like $X_i$ and $Y_i$ is not strictly local, in the sense that it affects a finite number of bonds.
Nevertheless, the measurements of $X_i$ and $Y_i$ are {\it quasi-local}, in the sense that their effect on measurements of observables at $i+l$ falls off exponentially quickly with the separation $l$.  
We also derive the expectation values in the canonical ensemble in this unusual Hilbert space. 
We restrict the discussion to infinite temperature; at infinite temperature, the exact form of $H$ plays no role and $\avg{\mathcal{O}}_{c}$ follows from the properties of random vectors in the constrained Hilbert space.

\subsection{Constraints and quasi-locality in the \goldchain Hilbert space} \label{CountingSec}

To quantify how far the constrained Hilbert space is from having a local product structure, we evaluate the dependence of measurements on bond $i$ on measurements at distant points $j$ in the chain -- i.e. we compare the probability
$
P(X_i = \tau , X_j = 1 ) 
$
to $P(X_i = \tau)P(X_j = 1)$.   

Consider the chain in Fig.~\ref{HFig1} with $(1,1)$ boundary conditions, i.e. with $X_0 = X_{N+1} =1$.
To evaluate $P(X_k = \tau)$, we partition the system into three pieces: subsystem $A$, which consists of the $k-1$ bonds to the left, bond $k$, and subsystem $B$ consisting of the $N-k$ bonds to the right.  
Let $n_A$ and $n_B$ respectively be the number of configurations in subsystems $A$ and $B$ with $(1,1)$ boundary conditions.  
Then:
\ba \label{xkEq}
x_{k} = \tau \Rightarrow n_A = n_{k} \ , \ \ n_B = n_{N-k+1} \n
x_{k} =1 \Rightarrow n_A = n_{k-1} \ , \ \ n_B = n_{N-k} 
\ea
where $n_m$ denotes the number of configurations of a $m$-bond chain with $(1,1)$ boundary conditions.
From Eq.~\eqref{HSize}:
\begin{align}
\label{Eq:Numbermbonds}
n_m = f_m
\end{align}
Eq.~(\ref{xkEq}) follows from the fact that if $x_{k} =1$, then both subsystems effectively have $(1, 1)$ boundary conditions.
If $x_{k} = \tau$, on the other hand, $x_{k\pm 1} \in \{1, \tau \}$, and each subsystem has the same number of configurations as a system with one additional bond and $(1,1)$ boundary conditions.  
We therefore obtain:
\ba \label{PXtau}
P(X_{k} = \tau) = \frac{n_{k}  n_{N-k +1}}{n_N} = \frac{ f_{k} f_{N-k+1} }{f_{N } } \n
P(X_{k} =1) = \frac{n_{k-1}  n_{N-k}}{n_N} = \frac{ f_{k-1} f_{N-k} }{f_{N } }
\ea

Using a similar reasoning, we can show that the random variables $X_k, X_{k+l}$ become independent as $l \to \infty$, with a correction exponentially small in $l$ for finite $l$.
Specifically, consider the joint probability $P(X_k = a, X_{k+l}=b)$. In this case, we divide our system into the two spins $X_k, X_{k+l}$ as well as three additional subsystems $A, B, C$; the joint probability $P(X_k =a, X_{k+l}=b)$ is given by counting the total number of configurations for each of the 3 subsystems, given that $X_k =a$ and $X_{k+l} =b$.  The result is:
\begin{align}
P(X_k = 1, X_{k+l}=1) &= \frac{f_{k-1}f_{l-1} f_{N-k-l}}{f_N} \n
P(X_k = \tau, X_{k+l}=1) &= \frac{f_{k}f_{l} f_{N-k-l}}{f_N}
\end{align}
and so on.
Thus, for example,
\begin{align}
&P(X_k = \tau, X_{k+l}=1)  - P(X_k = \tau) P(X_{k+l}=1)  \n
&=\frac{f_{k} f_{N-k-l}}{f_N} \left ( f_{l}  - \frac{f_{N-k+1} f_{k+l-1}}{f_N} \right)
\end{align}
We can estimate this difference using Binet's \cite{simon13} formula:
\be
f_n = \frac{\gmean^n - \psi^n}{\gmean- \psi } \  \ \text{ where } \ \ \psi \equiv  \frac{-1}{\gmean}
\ee
which gives:
\begin{align}
P(X_k = \tau, X_{k+l}=1) & - P(X_k = \tau) P(X_{k+l}=1) \n
&= (-1)^{l+1} \frac{\gmean^{-2 l}}{( \gmean- \psi)^2} \left( 1+  ... % - (-1)^k \varphi^{-2(k+1) }- (-1)^{N-k-l} \varphi^{-2(N -k-l +1)} - (-1)^n \varphi^{-N} 
\right )
\end{align}
where $...$ indicates terms at most on the order of min$(\gmean^{-2(k+1) }, \gmean^{-2(N -k-l +1)})$.
Thus the difference between the joint probability and the product of the two individual probabilities falls off exponentially in the separation $l$ between the two bonds. 
It is straightforward to check that the same holds for $P(X_k=1, X_{k+l}=1)$ etc, as well as for the local observables $Y_k$.  

We conclude that though the Hilbert space does not admit a local product structure, there remains a meaningful sense in which ``local" measurements can be made, since measurements of bonds separated by several times the correlation length $\xi = 1/(2\log(\gmean)) \approx 1.03$ are effectively independent.

\subsection{Expectation values in the canonical ensemble} \label{CanonicalSec}

To test ETH, we are interested in two quantities: the thermal expectation values of operators, and how quickly the distribution of eigenstate expectation values narrows with increasing system size. 
The infinite temperature expectation value of any operator is simply its Hilbert space average.  
From Eq.~(\ref{PXtau}):
\begin{align}
\label{Eq:ExpInfTX}
\avg{\Pi_{i;x}^1}_c = \frac{f_{i-1}f_{N-i}}{f_N}
\end{align}
To obtain $\avg{\Pi_{i;2}^1}_c$,  we need to evaluate the probabilities of the 5 possible configurations $(X_{i-1}, X_i, X_{i+1})$ of three consecutive bonds.
Using logic similar to that of Sec.~\ref{CountingSec}, we obtain the Hilbert space probabilities:
\ba \label{3BondProbs}
P(\tau,\tau,\tau) = P(\tau, 1,\tau)=&  \frac{ f_{N-i} f_{i-1} }{f_{N } }  \n
P(1, \tau, \tau) =& \frac{ f_{N-i} f_{i-2} }{f_{N } }  \n
P(\tau, \tau, 1) =& \frac{ f_{N-i-1} f_{i-1} }{f_{N } }  \n
P(1, \tau, 1 ) =&  \frac{ f_{N-i-1} f_{i-2} }{f_{N } }. 
\ea
From Eq.~\eqref{YProj},
\begin{align} \label{Yth}
\avg{\Pi_{i;2}^1}_c &= P(1, \tau, 1) + \gmean^{-1} P(\tau, 1, \tau) + \gmean^{-2} P(\tau, \tau, \tau) \nonumber\\
&\approx \gmean^{-2} +  O(\gmean^{-2N})
\end{align}
Eq.~(\ref{Yth}) gives the (bond-independent) $T = \infty$ thermal expectation value $\avg{\Pi_{i;2}^1}_c$.

Observe that {\it if} the bond variables $X_i$ thermalize, then local variables such as $\Pi_{i;2}^1$ also thermalize.   
This is because of the local change of basis between the triples $(X_{i-1},X_i, X_{i+1})$ and $Y_i$.  
If $\Pi^1_{i;x}$ thermalizes, then these triples should also thermalize --  indeed $X_i = 1$ implies $(X_{i-1},X_i, X_{i+1}) = (\tau,1,\tau)$, such that for one of the triples this is automatic.  
The argument does not work in reverse, however, since the bond variable $X_i$ cannot be reconstructed from the values of the $Y_j$ variables on bonds $j$ near bond $i$.  

A complementary perspective on Eq.~\eqref{Eq:ETHansatz} to that presented in Sec.~\ref{Sec:ETHIntro} is that expectation values of local observables in an ensemble of eigenvectors of the appropriate energy density behave exactly as they would in an ensemble of random eigenvectors in the same Hilbert space.  
In Appendix~\ref{RandomApp}, we verify explicitly that for an ensemble of random vectors in the $X$ basis, Eq.~(\ref{Eq:ETHansatz}) at infinite temperature holds both for $\Pi_{i;x}^1$ and $\Pi_{i;2}^1$.
This further demonstrates that if eigenvectors appear random in the bond ($X$) basis, then $Y$ (together with other local observables) appear thermal. 

Finally, we consider the braid operator between distant anyons defined in Eq.~(\ref{LongBraid}). 
In the infinite temperature ensemble, we expect that the average modulus of $B_{i, i+l}$ falls off exponentially in $l$, since the probability that all bonds are in the same state before and after braiding decreases exponentially with separation.
Further, the standard deviation $\Delta B_{i, i+l}$ decreases as the inverse square root of the number of configurations that yield a particular value of $B_{i, i+l}$.  
For fixed finite $l$, this number grows with the Hilbert space dimension of the chain $\varphi^N$.
If however $l/N$ is held fixed as $N\to\infty$, the number of configurations grows as $\varphi^{\alpha N}$ for some $\alpha\leq1$ (here $\alpha$ is at least as large as the fraction of the chain where no $R_{i, i+1}$ operators act).  
If $\alpha<1$, the variance of this operator decreases exponentially with $N$ at a slower rate than the local observables.

\section{Numerical studies}
\label{Sec:Numerics}
In this section, we investigate several properties of the highly excited eigenstates of the Hamiltonian in Eq.~\eqref{HChain} to test if ETH is obeyed.
We present data obtained by exact diagonalization at $\theta_e = \pi/4, \theta_o=\pi/5$ up to $N=23$ bonds (see Fig.~\ref{GCFig} for our labelling convention).
We have checked that the conclusions also hold at other values of $\theta_e, \theta_o$ away from integrable points.  The bulk of our study focuses on the observables $\Pi^1_{i; \alpha}$ for $i=N/2, N/2+1$ and $\alpha = x,2,3$, where for odd values of $N$, the site/bond label $N/2$ is understood to be rounded up to the nearest integer.  At the end of this section we briefly discuss the non-local braid operator $B_{i, i+l}$.

\begin{figure}[tbp]
\begin{center}
\includegraphics[width=\columnwidth]{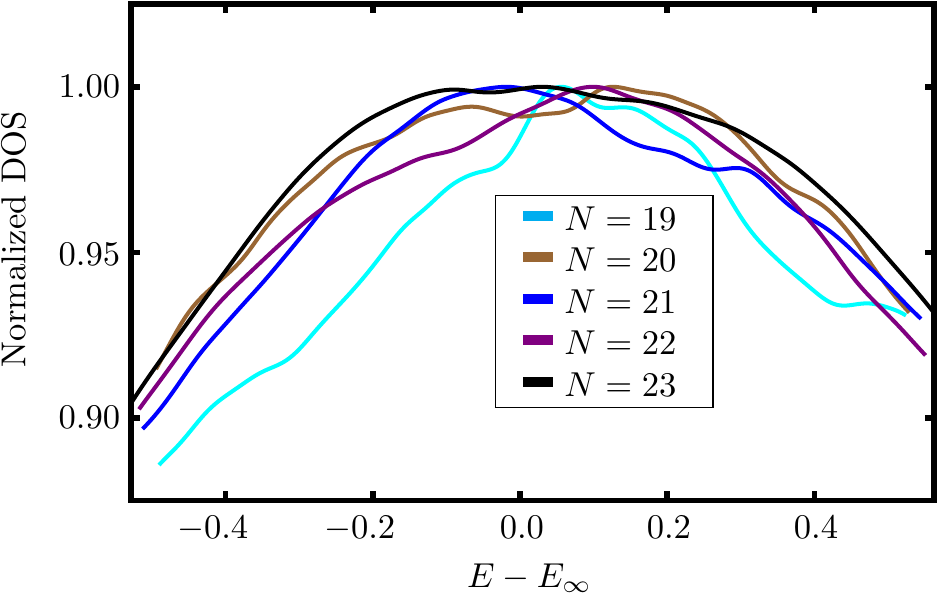}
\caption{The density of states (DOS) normalized by its maximum value as a function of energy for the five largest system sizes $N$.  Here $E_{\infty}=\textrm{Tr}{H}/\textrm{Tr 1}$ is the energy corresponding to infinite temperature in the canonical ensemble.  The normalized density of states is approximately constant over the middle $1/3$ of the spectrum.}
\label{Fig:ManybodyDOS}
\end{center}
\end{figure}

We restrict our attention to infinite temperature where the entropy and thus the number of states with fixed energy density is maximal at given $N$.  
As seen in Fig.~\ref{Fig:ManybodyDOS}, the many-body density of states (normalized by its maximum value) varies by less than $10\%$ in the center third of the spectrum.
This holds for system sizes $N\geq 14$ (not all shown).
It is therefore justified to treat the states in this range as infinite temperature states; below, $[\cdot]$ denotes averaging with respect to these states.

We begin with the statistics of the level spacings in the system.
This can be quantified using the level statistics ratio $r_m$, defined as \cite{Oganesyan:2007aa, atas13}:
\begin{align}
\label{Eq:RRatioDef}
r_m &\equiv \textrm{min} \left( \frac{\Delta E_{m}}{\Delta E_{m-1}}, \frac{\Delta E_{m-1}}{\Delta E_{m}}\right) \\
\Delta E_m &\equiv E_{m+1} - E_m.
\end{align}
Above $\Delta E_m$ is the $m$th level spacing and the energies are ordered $E_1 \leq E_2 \leq \ldots $, so that $r_m$ is the ratio of adjacent level spacings.
By definition, $0\leq r_m \leq 1$.

The level statistics ratio measures energy level repulsion in the spectrum. 
At integrable points or in localized systems, the energies are Poisson distributed and exhibit no level repulsion.
Consequently, the distribution of $r_m$ across the spectrum has non-zero density at $r_m=0$ with mean $[r_m] \approx 0.386$.
On the other hand, in thermalizing systems, we expect the distribution of eigenvalues to be given by random matrix theory (Wigner's surmise).
For our real Hamiltonian this implies that the level statistics should be that of the Gaussian Orthogonal Ensemble (GOE), with zero density at $r_m=0$ and a larger mean $[r_m] \approx 0.530$. 

\begin{figure}[tbp]
\begin{center}
\includegraphics[width=\columnwidth]{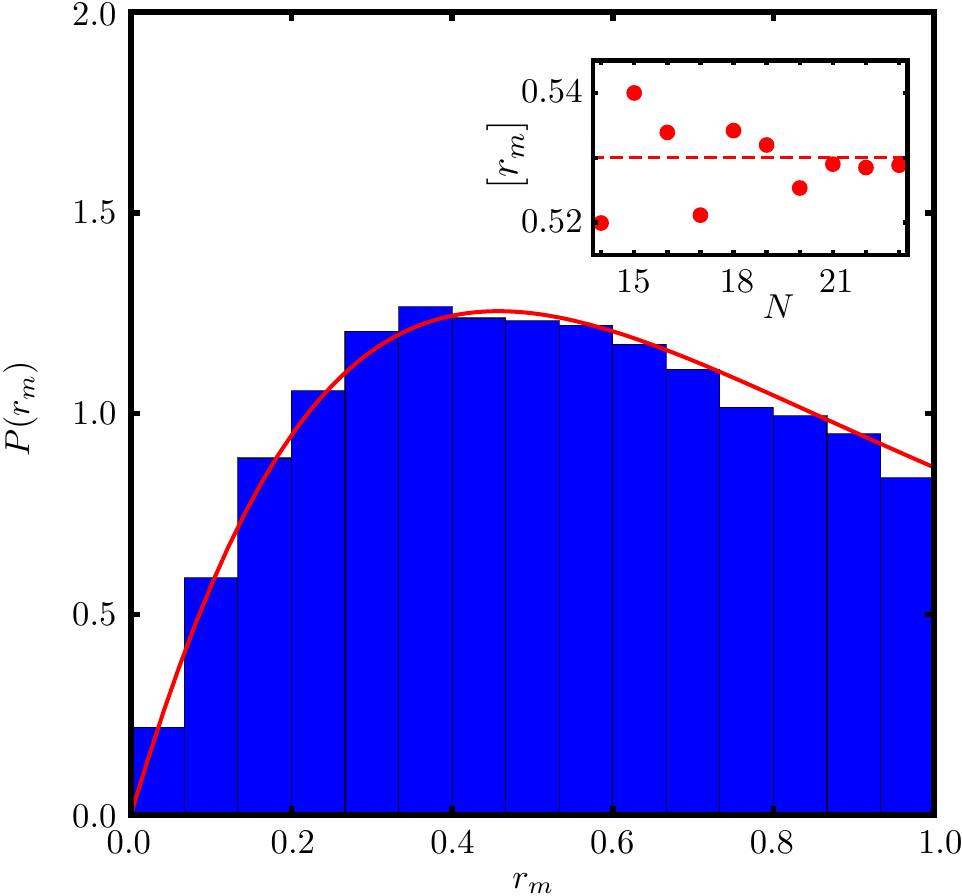}
\caption{Level statistics ratio $r_m$ (see Eq.~(\ref{Eq:RRatioDef})) for the Hamiltonian (\ref{HChain}).  The main figure shows the probability distribution of $r_m$ at the largest system size $N=23$ studied here; the red line shows the predicted distribution from Ref.~\onlinecite{atas13} for GOE statistics.  Inset: Mean of the level statistics ratio as a function of $N$ (red dots) compared to the mean of the ideal GOE distribution $\approx 0.530$ (dashed line). }
\label{Fig:RRatio}
\end{center}
\end{figure}

Shown in Fig.~\ref{Fig:RRatio} is the histogram of $r_m$ for the largest system size $N=23$.  
Qualitatively, the distribution shows the features of the thermalizing system.
It also agrees well with the Wigner-like surmise for the GOE ensemble from Ref.~\cite{atas13} (red curve).
The inset shows the mean value $[r_m]$ as a function of system size $N$ (dots) and the theoretical value of $0.530$ (line) for comparison.
The difference between the two is small, and clearly decreasing with $N$.

\subsection{Diagonal ETH for local and bond observables}

First, we investigate whether the observables $\Pi^1_{i; \alpha}$ for $i=N/2, N/2+1$ and $\alpha = x,2,3$ obey the diagonal ETH ansatz.  
In other words, we verify that (1) on average, the expectation value in an ensemble of eigenstates with a given energy density is that of the appropriate thermal ensemble, and (2) the fluctuations between eigenstates vanish exponentially with $N$ as $\exp(-S(E_n)/2)$, where $S(E_n)$ is the thermal entropy.

\begin{figure}[tbp]
\begin{center}
\includegraphics[width=\columnwidth]{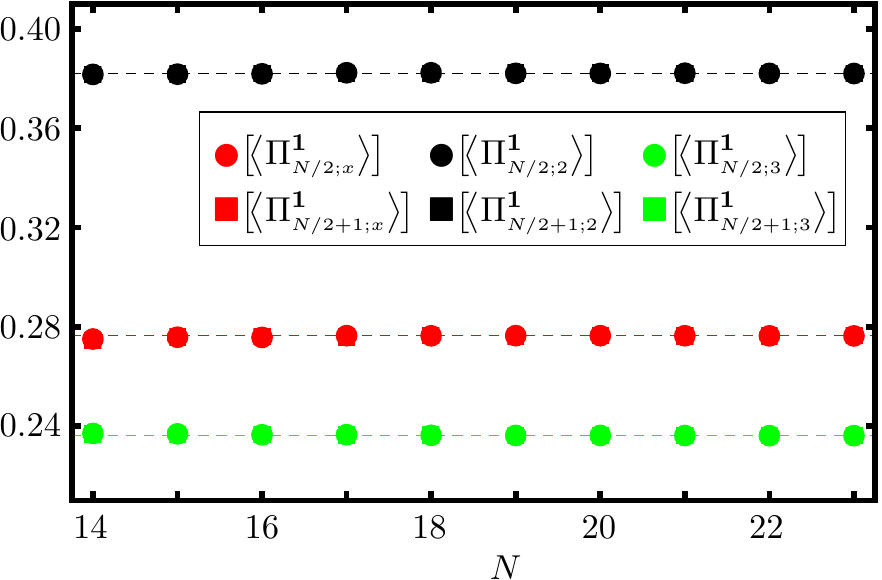}
\caption{Spectrum averaged expectation values of the operators $\Pi^1_{i;\alpha}$ for $i=N/2, N/2+1$ and $\alpha=x,2,3$. The dashed line in each case represents the predicted infinite temperature thermal value, which fits extremely well with the numerical data for all system sizes.}
\label{Fig:MeanLocalObs}
\end{center}
\end{figure}

In Fig.~\ref{Fig:MeanLocalObs}, we test (1).
The infinite temperature ensemble (dashed lines) and the eigenstate ensemble (points) are seen to reproduce the same answers for all six observables.
As expected, the dimerization in the Hamiltonian does not affect the value at infinite temperature, so that $[\langle \Pi^1_{N/2, \alpha} \rangle ] = [\langle \Pi^1_{N/2+1, \alpha}\rangle]$ for $\alpha = x,2,3$.

Next, we test (2) at infinite temperature, where $\exp(-S(E_n)/2) \sim \gmean^{-N/2}$.
Following Ref.~\cite{Kim:2014aa}, we define $\Delta \mathcal{O}_n$:
\begin{align} \label{DeltaOEq}
\Delta \mathcal{O}_n = \bra{n+1} \mathcal{O} \ket{n+1} - \bra{n} \mathcal{O} \ket{n}
\end{align}
$\Delta \mathcal{O}_n$ captures the fluctuations in Eq.~\eqref{Eq:ETHansatz} between adjacent eigenstates in the spectrum.
Its spectrum averaged absolute value $[ |\Delta \mathcal{O}| ]$ decreases exponentially as $\exp(-S(E_n)/2) $.  
The advantage of $\Delta \mathcal{O}_n$ over measures of the width of the distribution of $\langle n| \mathcal{O} |n \rangle$ in an energy window is that $\Delta \mathcal{O}_n$ is immune to the smooth variation of the mean value $[\bra{n}\mathcal{O}\ket{n}]$ in the energy window.

\begin{figure}[tbp]
\begin{center}
\includegraphics[width=\columnwidth]{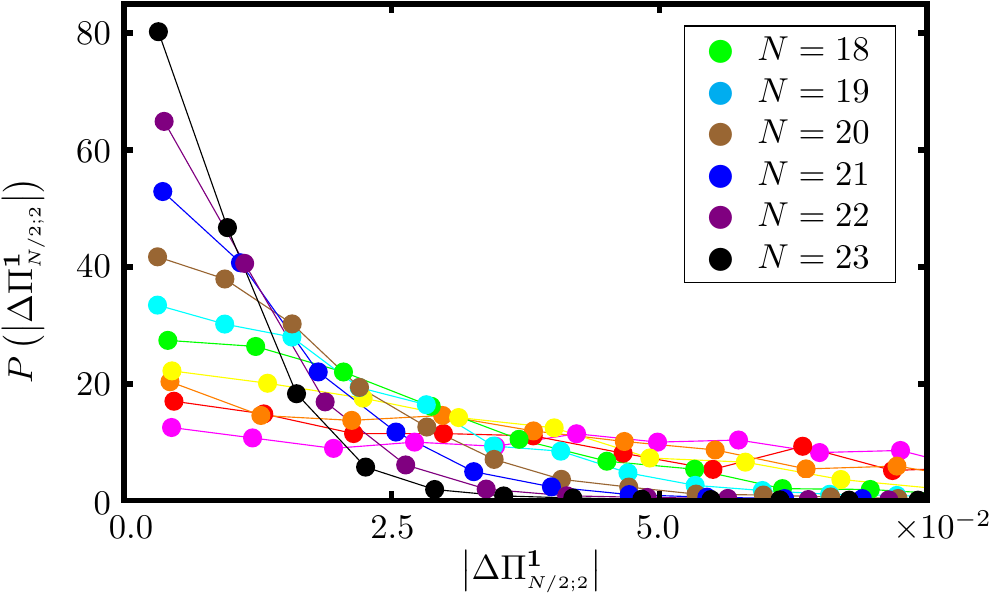}
\caption{Distribution of $|\Delta\Pi^1_{N/2; 2}|$ (see Eq.~(\ref{DeltaOEq})) for $N=18$ to $23$.  As predicted by ETH, the distribution becomes narrower and more sharply peaked about $0$ with increasing $N$. }
\label{Fig:PDistYDiff}
\end{center}
\end{figure}

\begin{figure}[tbp]
\begin{center}
\includegraphics[width=\columnwidth]{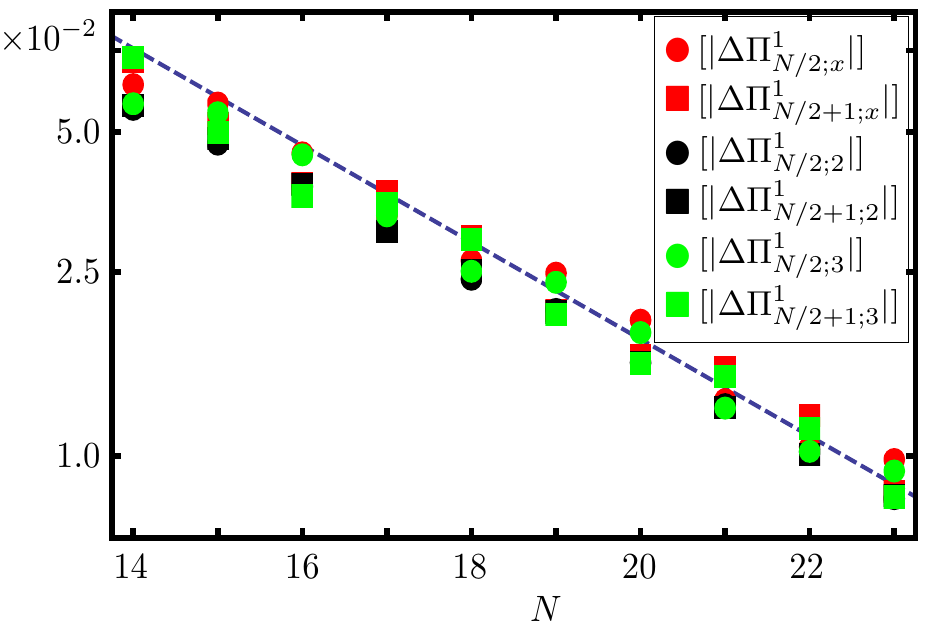}
\caption{Mean of $|\Delta \mathcal{O}|$ vs. the system size $N$ for different operators.  The dashed line indicates the predicted scaling behavior $[|\Delta \mathcal{O}|] \sim \varphi^{ - N/2}$, which is seen to be a good fit to the data.  Numerical least-squares best fit slopes of each operator are given in Table \ref{BSlopesTab}.} 
\label{Fig:MeanDiffLocalObs}
\end{center}
\end{figure}

Fig.~\ref{Fig:PDistYDiff} plots the probability distribution of $|\Delta \Pi^1_{N/2; 2}|$ at different $N$.
It is clearly seen that the distribution narrows with system size and peaks around zero, as expected for observables satisfying ETH.
In Fig.~\ref{Fig:MeanDiffLocalObs}, we plot $[|\Delta \mathcal{O}|]$ vs $N$ for the six observables on a log-linear plot.
The blue dashed line shows the predicted scaling proportional to $\varphi^{-N/2}$.
The fluctuations of the local and non-local observables decrease exponentially with $N$: the slopes of the best least-squares fit of $\log|\Delta \Pi^1_{N/2; 2}|$ vs $N$ are given in Table \ref{BSlopesTab}.   
The slope obtained from the largest system sizes is very close to the theoretical value of $- \frac{1}{2} \log \varphi \approx - 0.24$ for all observables measured.
Including the smaller system sizes in the fit reduces the absolute value of the slope as the effective ``temperature" (and therefore the entropy) shows larger variation in the center third of the spectrum at small $N$.

In summary, we have provided strong evidence that the Fibonacci chain satisfies diagonal ETH at infinite temperature.

\begin{table}[tbp]
\begin{center}
\begin{tabular}{|c|c|c|}
\hline
Variable & Best fit slope  & Best fit slope \\
& ($N=14$ to $N=23$) & ($N=19$ to $N=23$)\\
\hline
$ \Delta \Pi^1_{N/2; x}$ & $-0.221 \pm 0.009$ & $ -0.248\pm 0.032$\\
$ \Delta \Pi^1_{N/2+1; x}$ & $ -0.223\pm 0.009$ & $ -0.209\pm 0.031$\\
$ \Delta \Pi^1_{N/2; 2}$ & $ -0.217\pm 0.003$ & $ -0.232\pm 0.004$ \\
$ \Delta \Pi^1_{N/2+1;2}$ & $ -0.219\pm 0.002$ & $ -0.228\pm 0.005$  \\
$ \Delta \Pi^1_{N/2;3}$ & $ -0.220\pm 0.009$ & $ -0.246\pm 0.028$  \\
$ \Delta \Pi^1_{N/2+1;3}$ & $-0.225 \pm 0.010$ & $ -0.213\pm 0.026$  \\
\hline
\end{tabular}
\caption{Numerical least-squares best fit slopes for each of the operators shown in Fig.~\ref{Fig:MeanDiffLocalObs}.  To within the fitting error, for larger system sizes these agree well with the theoretical value of $- \frac{1}{2} \log \varphi \approx - 0.24$. }
\label{BSlopesTab}
\end{center}
\end{table}

\subsection{Off-diagonal ETH for local and bond observables} \label{Sec4B}
The ETH ansatz states that the off-diagonal matrix elements of a local operator $\mathcal{O}$ in the energy eigenbasis is distributed as:
\begin{align}
\bra{m} \mathcal{O} \ket{n} = \frac{1}{\sqrt{e^{S(\bar{E})}}} r_{mn} f_{\mathcal{O}}(\bar{E}, E_m - E_n)
\label{Eq:OffDiaETH}
\end{align}
where the symbols are explained below Eq.~\eqref{Eq:ETHansatz}.  In this section, we show that the off-diagonal matrix elements of local and bond observables in the Fibonacci chain obey Eq.~\eqref{Eq:OffDiaETH} at the average energy $\bar{E}$ corresponding to infinite temperature.

For this purpose, it is useful to note the following properties of the function $f_{\mathcal{O}}(E,\omega)$.  
First, $f_{\mathcal{O}}(E,\omega)$ is a smooth function of its arguments.  
Second, $f_{\mathcal{O}}(E,\omega) = f_{\mathcal{O}}(E,-\omega)$ (this follows from its definition in Eq.~\eqref{Eq:ETHansatz}).
Third, by Eq.~\eqref{Eq:fSpecDensity}, $f_{\mathcal{O}}(E,\omega)$ and the spectral density are proportional to one another.
Thus, the behavior of $f_{\mathcal{O}}(E,\omega)$ can be inferred from that of the spectral density.
If $\mathcal{O}$ involves a conserved density, then the spectral density (and consequently $f_{\mathcal{O}}$) diverges as $\omega \to 0$.
At finite size this divergence is cut off at small $\omega$ so that $f_{\mathcal{O}}(\bar{E}, \omega)$ is well approximated by a constant in a small energy window. 
At large $\omega$, both functions typically fall off exponentially with the energy difference $|\omega|$.

\begin{figure}[tbp]
\begin{center}
\includegraphics[width=\columnwidth]{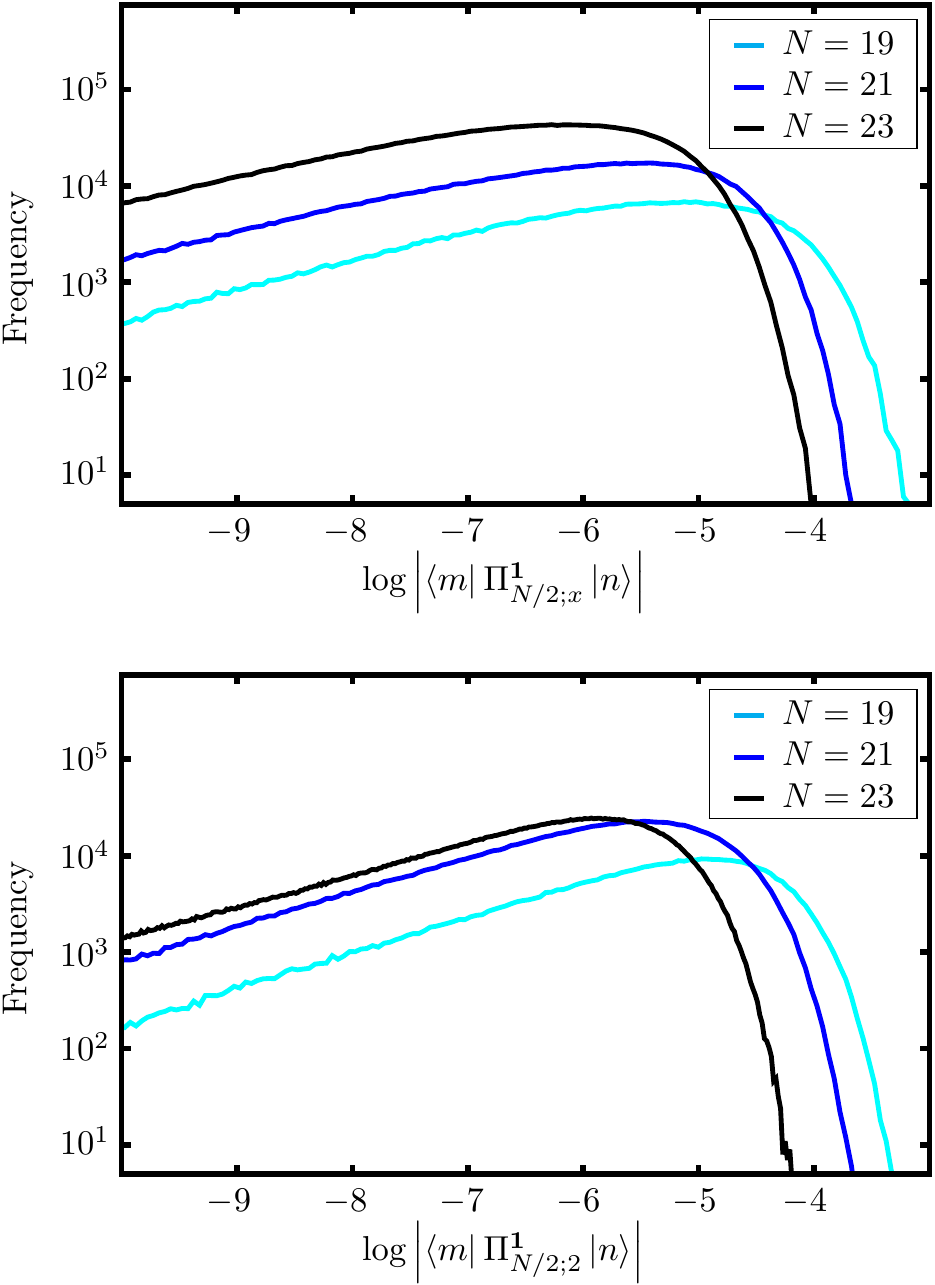}
\caption{ Histogram of $\log |\bra{n} \mathcal{O} \ket{n}|$ for $\mathcal{O} = \Pi^1_{N/2;x}$ (top) and $\mathcal{O} = \Pi^1_{N/2;2}$ (bottom) between hundred states at the center of the spectrum for $N=19, 21$ and $23$. The distribution shifts linearly to the left with increasing $N$ and the shape corresponds to the logarithm of the absolute value of a Gaussian distributed variable.  }
\label{Fig:MatElems}
\end{center}
\end{figure}

In Fig.~\ref{Fig:MatElems} is plotted the histogram of the logarithm of the absolute value of the matrix elements between a hundred states at infinite temperature at the three largest odd system sizes $N=19,21,23$.
The scale is log-linear.
The top panel is for the bond observable $\mathcal{O} = \Pi^1_{N/2;x}$, while the bottom panel is for the local observable $\mathcal{O} = \Pi^1_{N/2;2}$.
The first feature to notice is that the distribution of $\log|\bra{n} \mathcal{O} \ket{m}|$ rigidly shifts to the left with increasing $N$.
Indeed, when $m,n$ lie within a small energy window so that $f_{\mathcal{O}}(\bar{E}, E_m - E_n)$ is a constant $K$, Eq.~\eqref{Eq:OffDiaETH} predicts that at infinite temperature:
\begin{align}
\log|\bra{m} \mathcal{O} \ket{n}| = -\frac{N\log(\gmean)}{2} + \log|r_{mn}| + K
\label{Eq:OffDiagSmallWin}
\end{align}
Thus, ETH predicts that the distribution shifts to the left by $-\log(\gmean) \approx -0.48$ when $N$ is increased by two sites.
This is in good agreement with both panels of Fig.~\ref{Fig:MatElems} and with the behavior of the mean:
\begin{align}
 [\log|\bra{n} \Pi^1_{N/2;x}\ket{m}|] &\approx -0.48N + \textrm{constant} \\
 [\log|\bra{n} \Pi^1_{N/2;2}\ket{m}|] &\approx -0.43N + \textrm{constant}
 \end{align}
 The three-body local observable $\Pi^1_{N/2;3}$ exhibits a similar behavior.
 
The second term in Eq.~\eqref{Eq:OffDiagSmallWin} predicts the shape of the distribution to be the probability distribution of the logarithm of the absolute value of a Gaussian distributed random variable.
On defining $\log|r_{mn} | \equiv G = \log|\bra{m} \mathcal{O} \ket{n}| + N\log(\gmean)/2 - K $, a simple change of variables leads to:
\begin{align}
P(G=y) = \frac{2}{\sqrt{ 2\pi } } e^{y} \exp{\left(-e^{2y}/2\right)}.
\end{align} 
With the single fitting parameter $K$ for each observable, we checked that this function describes the shape of the full distribution at all $N$.
In particular, it explains the exponential decay seen in Fig.~\ref{Fig:MatElems} for $y < [\log|\bra{m} \mathcal{O} \ket{n}| ]$ and the steep double exponential decay for $y> [\log|\bra{m} \mathcal{O} \ket{n}| ]$.
Thus, the ETH prediction Eq.~\eqref{Eq:OffDiagSmallWin} holds for local and bond observables in the non-integrable Fibonacci chain.

A different way to test off-diagonal ETH and Eq.~\eqref{Eq:OffDiaETH} is using the inverse participation ratio (IPR) of $\mathcal{O}\ket{n}$ in the energy eigenbasis:
\begin{align} \label{IPRDef}
I^\mathcal{O}_n = \sum_m |\bra{m} \mathcal{O}\ket{n}|^4
\end{align}
The inverse of $I^\mathcal{O}_n$ is a measure of the number of energy eigenstates with weight in $\mathcal{O}\ket{n}$ (the participation ratio).
ETH predicts that that this number is maximal and set by the number of energy eigenstates $e^{S(E_n)}$ at the energy $E_n$.
Indeed, using Eq.~\eqref{Eq:OffDiaETH} and the properties of  $f_{\mathcal{O}}(E,\omega)$ listed above, it is easy to show that \cite{DAlessio:2015aa}:
\begin{align}
I^\mathcal{O}_n \sim \frac{1}{e^{S(E_n)}}
\label{Eq:IPRPredETH}
\end{align}

\begin{figure}[tbp]
\begin{center}
\includegraphics[width=\columnwidth]{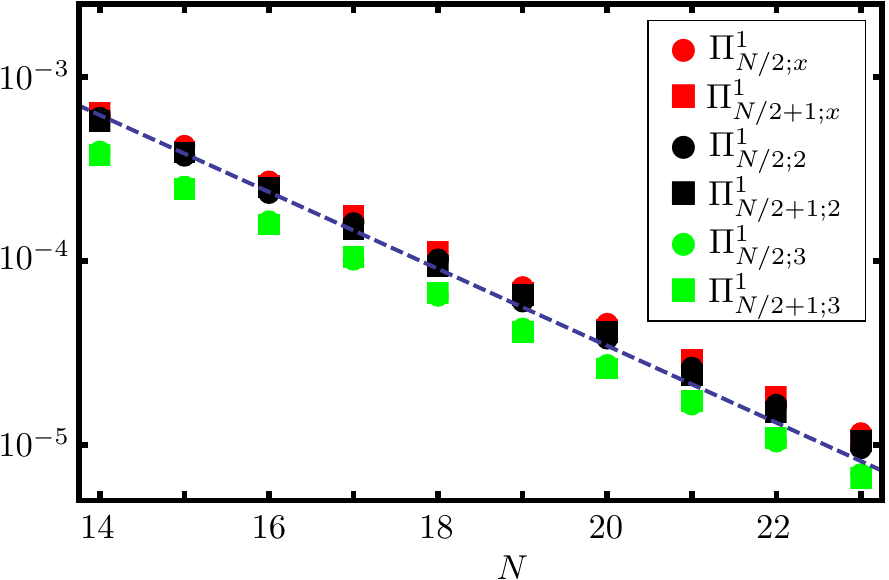}
\caption{Exponential of $[\log(I_n^\mathcal{O})]$ for the observables indicated in the legend.  The black dashed line indicates the predicted scaling behavior $I^\mac{O}_n \sim \varphi^{ - N}$, which is seen to be a good fit to the data.  Numerical least-squares best fit slopes of each operator are given in Table \ref{SlopesTab2}. }
\label{Fig:IPatRat}
\end{center}
\end{figure}

In Fig.~\ref{Fig:IPatRat} is plotted the exponential of the spectrum average of $\log(I_n^\mathcal{O})$ for the six observables $\mathcal{O}=\Pi^1_{i;\alpha}$, $i=N/2, N/2+1$ and $\alpha = x,2,3$ on a log-linear scale.
The dashed line is the ETH prediction Eq.~\eqref{Eq:IPRPredETH} at infinite temperature.
It is a straight line on the log-linear plot with slope $-\log(\gmean)$. 
The data points also fall on a straight line whose slope seems to approach $-\log(\gmean)$ with increasing $N$.
This is quantified by the numerical best fit slopes, listed in Table \ref{SlopesTab2}.

\begin{table}[tbp]
\begin{center}
\begin{tabular}{|c|c|c|}
\hline
Variable & Best fit slope  & Best fit slope \\
& ($N=14$ to $N=23$) & ($N=19$ to $N=23$)\\
\hline
$ \left[I^{ {N/2;x}} \right]$ & $ -0.450\pm 0.005$ & $ -0.465\pm 0.015$\\
$ \left[I^{{N/2+1;x}} \right]$ & $ -0.465\pm 0.015$ & $ -0.448\pm 0.005$\\
$ \left[I^{ {N/2;2}} \right]$ & $ -0.454\pm 0.005$ & $ -0.450\pm 0.018$  \\
$ \left[I^{ {N/2+1;2}} \right]$ & $ -0.453\pm 0.005$ & $ -0.465\pm 0.017$  \\
$ \left[I^{ {N/2;3}} \right]$ & $ -0.452\pm 0.005$ & $ -0.461\pm 0.009$  \\
$ \left[I^{ {N/2+1;3}} \right]$ & $ -0.448\pm 0.003$ & $-0.453 \pm 0.008$  \\
\hline
\end{tabular}
\caption{Numerical least-squares best fit slopes for each of the inverse participation ratios (see Eq.~(\ref{IPRDef})) shown in Fig.~\ref{Fig:IPatRat}. For notational brevity, we shorten the projector $\Pi^1_{i;\alpha}$ to simply $i;\alpha$ in the superscript. The slopes slightly underestimate the theoretical value of $-  \log \varphi \approx - 0.48$. }
\label{SlopesTab2}
\end{center}
\end{table}

\begin{figure}[tbp]
\begin{center}
\includegraphics[width=\columnwidth]{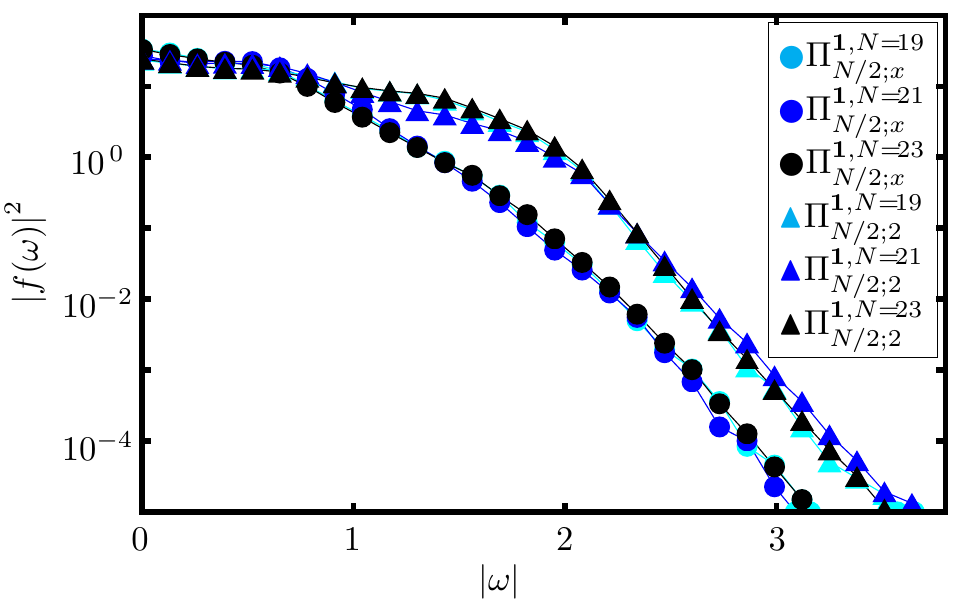}
\caption{ The function  $|f( \omega)|^2$ (see Eq.~\eqref{Eq:OffDiaETH}) as a function of  $|\omega|$ for $\mac{O} = \Pi^1_{N/2; x}$ (circles) and $\mac{O}=\Pi^1_{N/2; 2}$ (triangles) at mean energy $\bar{E}$ corresponding to infinite temperature.}
\label{Fig:ffunctionXY}
\end{center}
\end{figure}

Our final test of off-diagonal ETH consists of verifying the expected behavior of the smooth function $f_{\mathcal{O}}(\bar{E}, \omega)$ in Eq.~\eqref{Eq:OffDiaETH}.
Fig.~\ref{Fig:ffunctionXY} plots $|f( \omega)|^2 \equiv |f(\bar{E}, \omega)|^2$ at the mean energy $\bar{E}$ corresponding to infinite temperature versus $|\omega|$ for two different observables.
The colors indicate the system size $N$, while the circular (triangular) markers indicate the observable $\Pi^1_{N/2; x}$ ($\Pi^1_{N/2;2}$).
Note that we plot $|f( \omega)|^2$ vs $|\omega|$ as it is an even function of $\omega$.

First, observe that $|f( \omega)|^2$ decays exponentially on a scale $\Delta$ at large $|\omega|$ independent of $N$ for both observables.
From Eq.~\eqref{Eq:fSpecDensity}, $\Delta^{-1}$ therefore sets the time-scale for the local dynamics of the connected correlator $\bra{ n} \mathcal{O}(t)\mathcal{O}(0) \ket{n}_c$. 
Notice that $\Delta$ for the bond observable is less than that for the local observable $\Pi^1_{N/2;2}$.
This agrees with the expectation that local dynamics is less effective in relaxing the non-local bond observable  as compared to the observable that measures the local fusion channel.
At small $\omega$, $|f(\omega)|^2$ is expected to diverge in the thermodynamic limit, as both observables have some overlap with the conserved energy density. 
While the data at system sizes $N=19, 21, 23$ exhibits evidence of this divergence (for example, $|f(0)|^2$ increases with $N$), the data is too noisy to extract the functional form or the diffusion constant.

\subsection{Thermalization of non-local braid observables}

\begin{figure}[tbp]
\begin{center}
\includegraphics[width=\columnwidth]{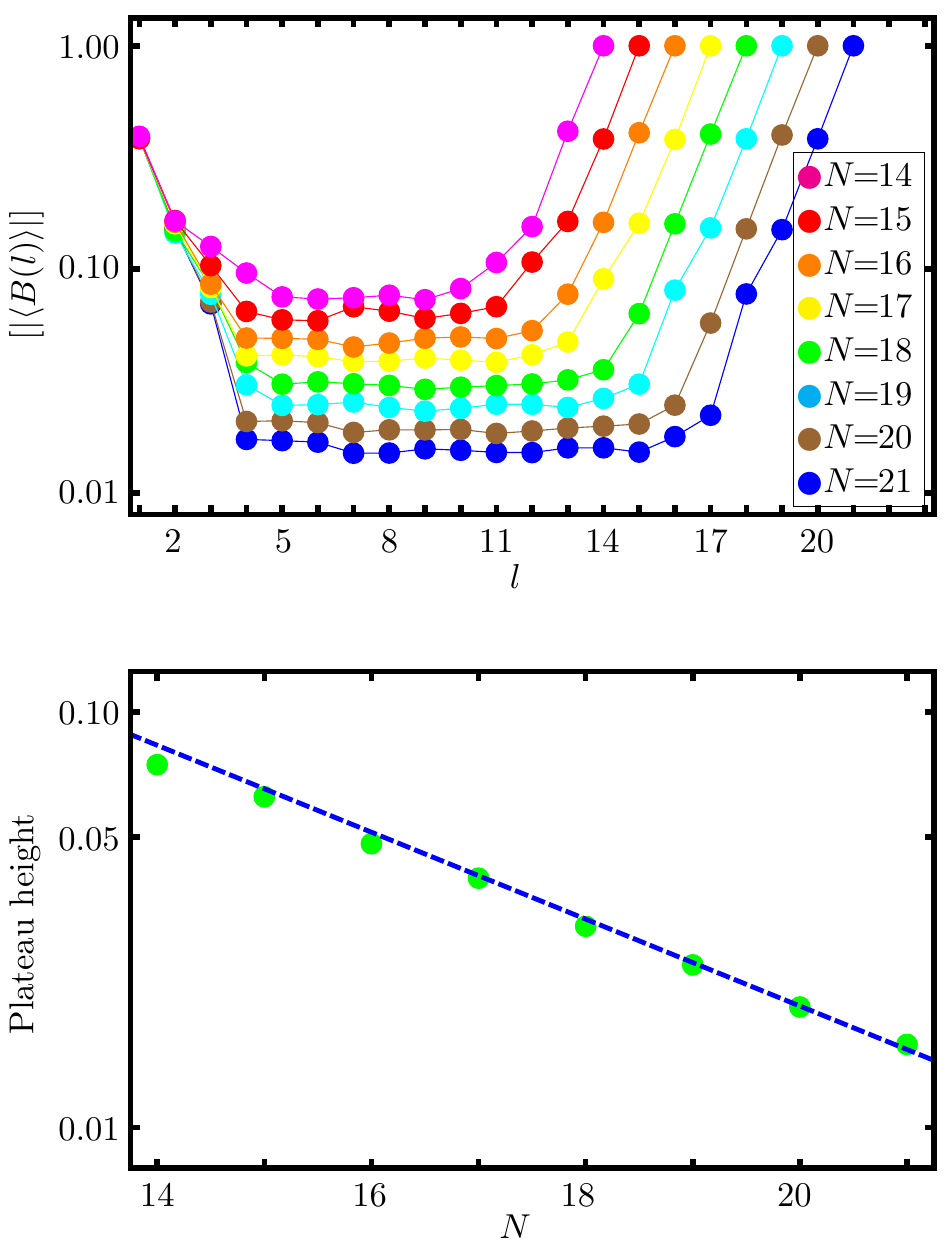}
\caption{ (a) Expectation of the modulus of the braid operator $[ |\langle B(l) \rangle |]$ as a function of separation $l$, for $N= 14$ to $21$.   The modulus falls off exponentially with $l$ for small $l$, reaching a plateau at approximately $l \geq 4$, and then exponentially increases as $l$ approaches $N$.   (b) The height of the plateau for $4 \leq l \leq N-5$ as a function of system size $N$.  This height falls off as $\gmean^{- N/2}$, as expected for a random positive quantity (dashed line).}
\label{BraidAmpFig}
\end{center}
\end{figure}

\begin{figure}[tbp]
\begin{center}
\includegraphics[width=\columnwidth]{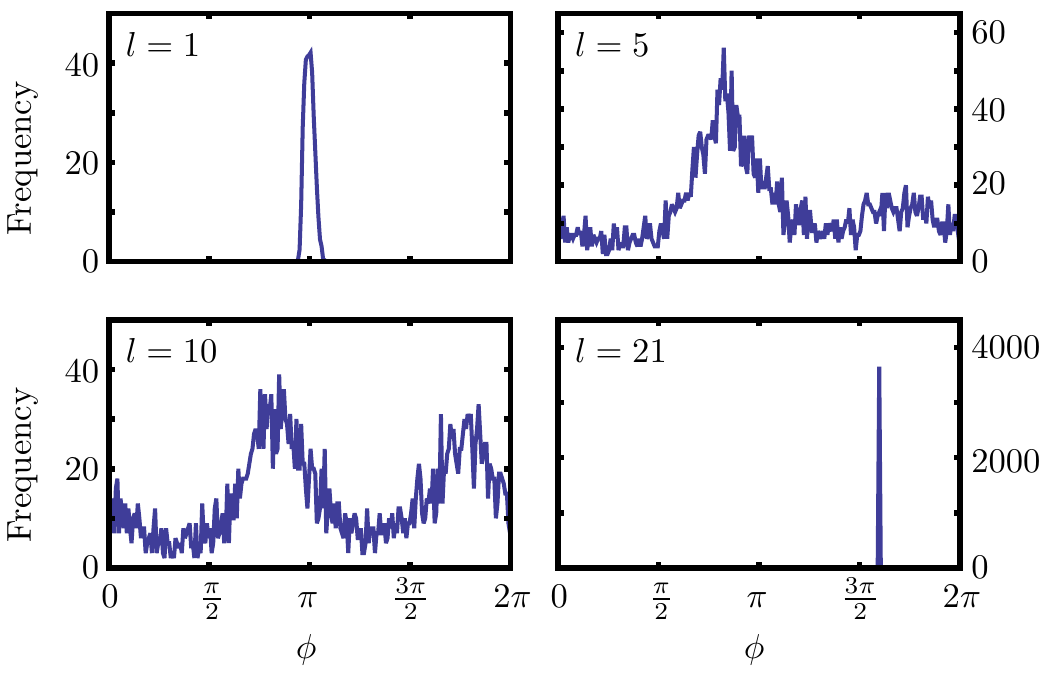}
\caption{ Histogram of phases $\phi$ of the braid operator $B(l)$ in the middle third of the eigenvectors as a function of separation $l$ for $N=21$.   For $l=1$, the phases are sharply peaked about $\pi$.  For $l= 5,10$, the phases are relatively uniformly distributed in the interval $( 0, 2\pi)$.  For $l=N$ the phase takes on only one value of $ 8 \pi/5 \approx 1.6 \pi$; this value is fixed by the boundary conditions. }
\label{BraidAmpFigHist}
\end{center}
\end{figure}

\begin{figure}[tbp]
\begin{center}
\includegraphics[width=\columnwidth]{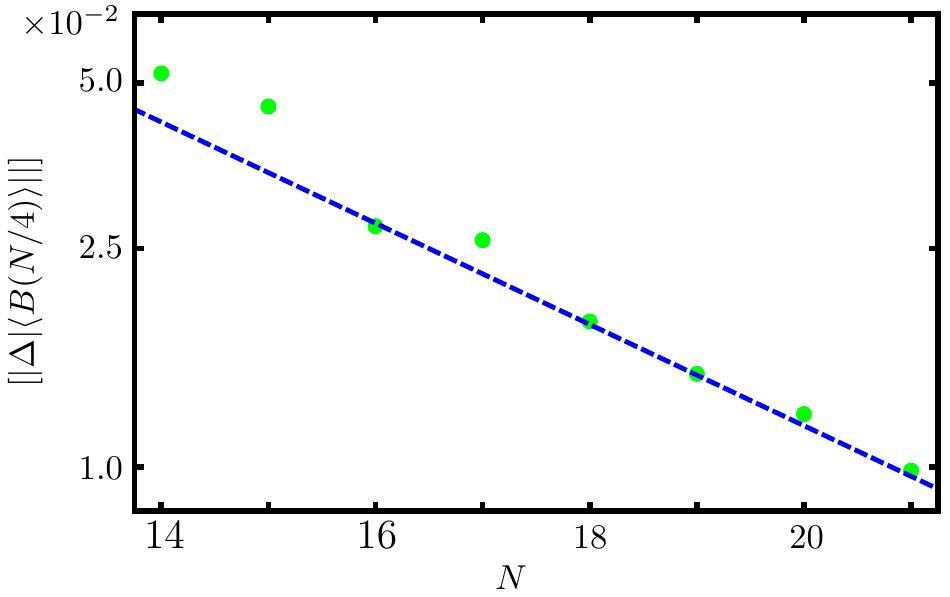}
\caption{ $[|\Delta |\langle B(N/4) \rangle ||]$ as a function of system size $N$.  The decrease is consistent with a scaling proportional to $\varphi^{-N/2}$ (dashed line), exactly as for local observables.}
\label{BraidDeltaFig}
\end{center}
\end{figure}

In this section, we show that even non-local operators like the braid operator $B_{i, i+l}$ satisfies diagonal ETH at infinite temperature. For notational brevity, define:
\begin{align}
B_{(N-l)/2+1, (N+l)/2+1}\equiv B(l)
\end{align}

As discussed in Sec.~\ref{CanonicalSec}, $[ |\langle B(l)\rangle |]$ decreases exponentially with $l$ in an ensemble of random states.  
This is indeed seen in Fig.~\ref{BraidAmpFig}; $ [|\langle B(l) \rangle|]$ initially decreases exponentially with $l$, rapidly reaching a plateau for $4 \leq l \leq N-5$.  
Further, as in an ensemble of random states, the height of this plateau decreases with system size, approximately as $1/\varphi^{N/2}$.

The figure also clearly shows that as $l$ approaches the system size, the modulus increases, reaching $1$ when the initial and terminal anyons are braided.  
This behavior is also expected for random states: as the net fusion channel of all anyons in the chain is fixed, for $l=N$ there is only one possible outcome for the braid.  Similarly for $l$ close to the total system size, the limited variance in the fusion channels reduces the number of different fusion channels that can occur in each eigenstate, leading to the observed increase in $[ |\langle B(l) \rangle|]$.

A similar trend is also visible in the distribution of braiding phases, shown for $N=21$ in Fig. \ref{BraidAmpFigHist}.  
Take $l=1$.
Using Eq.~\eqref{Eq:Rmatdef}, Eq.~\eqref{Yth} and the definition of $B(1)$, we obtain:
\begin{align*}
[\langle B(1) \rangle] &= P(Y=1) e^{ 8 \pi i/5} + P(Y=\tau) e^{ - 6 \pi i /5} \\
&= - \varphi^{-2} \approx - 0.38
\end{align*}
in an ensemble of random states.
The eigenstate ensemble reproduces the same values: the modulus of $B(1)$ is close to $0.38$ in Fig.~\ref{BraidAmpFig}, while the phase is sharply peaked about $\pi$.  
As $l$ increases -- such that the result of the braid depends on an ensemble of values of $Y_i$ --  the distribution of phases observed in the eigenstate ensemble broadens.  
This suggests that the ensemble of $Y_i$ is essentially random in a given eigenstate, and further indicates that no useful information is captured by these phases for most values of $l$.  
Again the exception is when an anyon is braided around almost the entire system: in this case, the net fusion channel of all anyons in the chain ($=1$) constrains the ensemble of values of $Y_i$ and the distribution of the phases is $\delta( \phi - 8 \pi /5 )$.

These measurements suggest that at infinite temperature, the braid operator $B(l)$ behaves as it should for a random state in the Hilbert space.
As a final check, in Fig.~\ref{BraidDeltaFig}, we plot $[|\Delta |\langle B(N/4)\rangle ||]$ vs $N$, where $[|\Delta |\langle B(N/4)\rangle ||]$ is the spectrum averaged absolute value of $\Delta |\langle B(N/4) \rangle|$ (see Eq.~\eqref{DeltaOEq}).
As anticipated, this quantity decreases exponentially with $N$.  
The exponent is approximately consistent with a fall-off of $\varphi^{-N/2}$, identical to that of local observables.

\section{Discussion } \label{Sec:DiscussionSection}
In this article, we have provided analytical and numerical evidence that non-integrable chains of pinned non-Abelian anyons obey the eigenstate thermalization hypothesis (ETH).
We argued that despite the lack of a strict tensor product structure to the Hilbert space of the non-Abelian anyons, both $\Pi^1_{i;x}$ and $\Pi^1_{i;2}$ are ``exponentially local", in the sense that the interdependence of measurements on different bonds falls off exponentially with the separation.  
As the notion of locality is not destroyed by the local constraints, this suggested that ETH holds in non-Abelian systems.
We then checked this claim numerically in a non-integrable model of Fibonacci anyons in 1D.
We showed that the expectation values of local observables (like the projection onto the trivial channel for few nearby anyons) and non-local observables (like the projection onto the trivial channel for half the anyons) in individual eigenstates coincided with the thermal value.
The fluctuations between eigenstates decreased exponentially as $\gmean^{-N}$.
We also showed that the off-diagonal matrix elements of local and non-local observables in the energy eigenbasis is a random matrix when the energy difference is small and extracted the smooth energy dependence at larger energy differences.
As the ETH ansatz (including the off-diagonal structure) guarantees thermalization in a dynamical experiment, we conclude that the Fibonacci chain acts as its own thermal reservoir in isolation.

Additionally, we established that at infinite temperature, non-local operators like braiding an anyon around a fraction of the total number of anyons also behaves as expected for random eigenvectors.  In general this implies that, at high temperatures, such non-local braids do not have any special properties facilitating information storage.  The notable exception occurs when the anyons braided are the first and last anyons in the chain, in which case the outcome of the braiding process is uniquely fixed by the boundary conditions.

Though the present work focuses specifically on the Fibonacci chain, we expect the conclusions to apply in any dimension to other pinned anyon models and more generally, to any lattice gauge theory. All these models come with local constraints which guarantee a notion of locality in the constrained Hilbert space (as made precise in Sec.~\ref{Sec:ThermalizationFib}). 
We therefore expect that they satisfy ETH and thermalize under their own dynamics.

As a specific example, consider the following generalization of the Fibonacci chain to 2D.
The pinned $\tau$ anyons form a honeycomb lattice in which each edge can be labeled either $1$ or $\tau$, with the constraint that the three legs entering each vertex must fuse to $\tau$.  
This model has an Ising-type Hilbert space, subject to the local constraint that no vertex can be surrounded by three $1$-type edges.  
More generally, any lattice gauge theory (or string-net model \cite{Levin:2005bs}) with non-dynamical matter fields necessarily obeys a similar constraint at each vertex, ensuring that the (fixed) matter field at that vertex sources the net electric flux leaving the vertex.  
Our results suggest that all such models obey ETH.

It would be interesting to test these expectations in general anyon models and in 2D, and in particular to explore thermalization in itinerant anyons, as discussed for example in Refs.~\cite{poilblanc13b,soni16}.
We note that the fusion tree basis obtained by enumerating the anyons and ordering them into a line is not illuminating for the question of thermalization in $2$D as the Hamiltonian appears non-local in this basis. 

In conclusion, integrability seems to be the only impediment to thermalization  in isolated quantum systems.
When the system is clean, the Hamiltonian must be fine-tuned to make the system integrable, resulting in an extensive number of conservation laws.  Such systems thermalize to a generalized Gibbs ensemble (GGE) with extra chemical potentials to account for the extra conservation laws \cite{Jaynes:1957aa,Jaynes:1957ab,Rigol:2007cr}.
Recent experiments in cold gases have quantitatively tested the effects of the GGE in 1D \cite{Trotzky:2012uq,Langen:2013aa,Langen:2015aa}.
Strongly disordered systems can also generically fail to thermalize in the thermodynamic limit and violate ETH if they are many-body localized \cite{Anderson:1958ly,Basko:2006aa,Oganesyan:2007aa,Pal:2010gs,Chandran:2016aa}.
Integrability underlies this failure (at least in strongly disordered regimes \cite{Huse:2014aa,Serbyn:2013rt}), although there is no GGE description of the steady state.  
Our work shows that, unlike integrability, local constraints and non-Abelian statistics do not impede thermalization.

\begin{acknowledgements}
The authors are grateful to David Huse, Chris Laumann and Shivaji Sondhi for helpful discussions and comments on a draft of this article, and to Siddharth Parameswaran for discussions. FJB and MDS are thankful to the Perimeter Institute for hospitality during the course of this work. FJB is supported by NSF DMR-1352271 and Sloan FG-2015-65927. Research at the Perimeter Institute is supported by the Government of Canada through Industry Canada and by the Province of Ontario through the Ministry of Economic Development and Innovation.
\end{acknowledgements} 

\appendix

\section{3-body interaction terms} \label{3BodyApp}

The 3-body interaction described in the main text is obtained in two steps.  First, we perform the two subsequent basis transformations shown in Fig.~\ref{HFig1}(b), to obtain the variable $Z_i$.  In this case, unlike the single basis transformation used to obtain $Y_i$, after the first step the left-most vertical leg can now be in state $1$ or $\tau$, leading to a slightly different matrix structure.   The resulting net basis transformation from the 8 allowed quadruples 
$(X_{i-1}, X_{i}, X_{i+1}, X_{i+2} ) =   (1, \tau, \tau,1)$, $(1, \tau, 1, \tau)$,  $(1, \tau, \tau,\tau)$,  $ (\tau, 1, \tau, 1)$, $(\tau, \tau, \tau,1)$, $ (\tau, \tau, 1, \tau)$,   $(\tau, 1, \tau, \tau)$, and $(\tau, \tau, \tau,\tau)$ onto the $8$ possible states 
$(X_{i-1}, Y_{i}, Z_i, X_{i+2} ) = (1, \tau, 1, 1)$,  $ (1, 1, \tau, \tau)$, $ (1, \tau, \tau, \tau)$, $(\tau, 1,\tau, 1)$, $(\tau, \tau,\tau, 1)$, $ (\tau, \tau, 1,\tau)$,  $ (\tau, 1, \tau, \tau)$, and $(\tau, \tau, \tau,\tau)$ is given by
\begin{align*}
\left(
\begin{array}{cccccccc}
 1 & 0 & 0 & 0 & 0 & 0 & 0 & 0 \\
 0 & 1 & 0 & 0 & 0 & 0 & 0 & 0 \\
 0 & 0 & 1 & 0 & 0 & 0 & 0 & 0 \\
 0 & 0 & 0 & \gmean^{-1} & \gmean^{-1/2} & 0 & 0 & 0 \\
 0 & 0 & 0 & \gmean^{-1/2} & -\gmean^{-1} & 0 & 0 & 0 \\
 0 & 0 & 0 & 0 & 0 & \gmean^{-1} & \gmean^{-1} & -\gmean^{-3/2} \\
 0 & 0 & 0 & 0 & 0 & 0 & \gmean^{-1} & \gmean^{-1/2} \\
 0 & 0 & 0 & 0 & 0 & \gmean^{-1/2} & -\gmean^{-3/2} & \gmean^{-2} \\
\end{array}
\right)
\end{align*}
Second, the projector onto states with $Z_i=1$ can be expressed in terms of quadruples of the $X_i$ by taking this basis transformation, projecting onto states with $Z_i=1$, and then inverting the result.  This gives
\begin{align*}
\Pi^1_{i;3} =& \n
&\left(
\begin{array}{cccccccc}
 1 & 0 & 0 & 0 & 0 & 0 & 0 & 0 \\
 0 & 0 & 0 & 0 & 0 & 0 & 0 & 0 \\
 0 & 0 & 0 & 0 & 0 & 0 & 0 & 0 \\
 0 & 0 & 0 & 0 & 0 & 0 & 0 & 0 \\
 0 & 0 & 0 & 0 & 0 & 0 & 0 & 0 \\
 0 & 0 & 0 & 0 & 0 & \gmean^{-2} & \gmean^{-2} & -\gmean^{-5/2} \\
 0 & 0 & 0 & 0 & 0 & \gmean^{-2} & \gmean^{-2} & -\gmean^{-5/2} \\
 0 & 0 & 0 & 0 & 0 & -\gmean^{-5/2} & -\gmean^{-5/2} & \gmean^{-3} \\
\end{array}
\right)
\end{align*}

\section{ Hamiltonians with potential inversion symmetry} \label{App:InversionSymm}

Here we briefly discuss data for non-dimerized chains $\theta_e= \theta_o =\pi/4$.  
The GOE level statistics can be recovered in this case by sorting eigenvectors according to their parity eigenvalues, and keeping only eigenvectors of parity eigenvalue $+1$.  In principle this is an advantage, as the Hamiltonian could be constructed explicitly in the $P = +1$ eigenbasis, allowing slightly larger system sizes to be achieved.  In practice, however, the variance of observables near the center of the chain in the parity-even Hilbert space shows a strong even-odd effect even for random vectors, making the data somewhat more difficult to interpret.  For this reason in the main text we have presented results where parity is explicitly and strongly broken.

The even-odd effect for bonds near the centre of the chain arises from an even-odd modulation in the fraction of states with a particular value of $X$ that have even parity.  
Consider the states with $X_{N/2}=1$.
If $N$ is even, then parity interchanges the two adjacent bonds: $N/2+1$ and $ N/2$.  However, configurations with $X_{N/2+1}= X_{N/2} =1$ are not allowed in the \goldchain Hilbert space.  
Thus, $X_{N/2} = 1$ implies $X_{N/2+1} = \tau$, and the configurations on the left and right halves of the chain are never identical.
Consequently, for every even parity state, there is an odd parity state and the number in each parity sector is half the total number of configurations in the Hilbert space with $X_{N/2} =1$.
If $N$ is odd, on the other hand, then (using our convention that $N/2$ is rounded up to the nearest integer) parity takes bond $N/2$ to itself. 
In this case there are clearly more states with $X_{N/2} = 1$ that have even parity than have odd parity, since there are $f_{N/2 -1}$ such bond configurations which are invariant under parity.  

\begin{figure}[tbp]
\begin{center}
\includegraphics[width=\columnwidth]{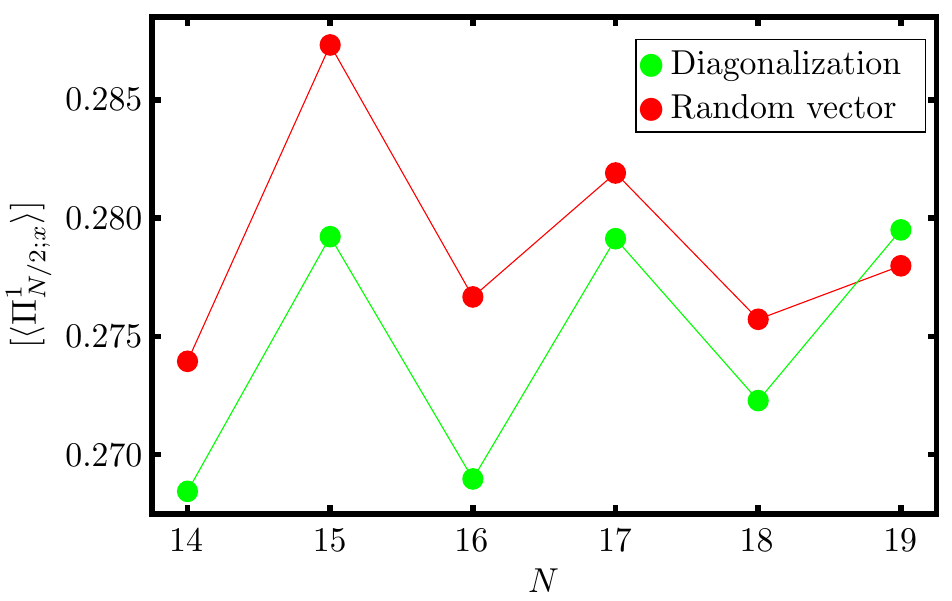}
\caption{$[\langle \Pi^1_{N/2, x} \rangle]
$ for 500 parity even random vectors (green, dashed line) and the central third of eigenvectors (solid, red line), as a function of system size $N$.  Restricting to even parity gives a larger fraction of states with $\Pi^1_{N/2, x} =1$ for $N$ odd than for $N$ even, as explained in the text; this is reflected in the larger value of $\langle \Pi^1_{N/2, x} \rangle$ for $N$ odd in both eigenvectors and random vectors. }
\label{Fig:RandEO}
\end{center}
\end{figure}

In other words, for even $N$ exactly half the configurations with $X_{N/2}=1$ have even parity, while for odd $N$ the corresponding fraction is greater than half.
As the fraction of configurations with $X_{N/2}=1$ is the expectation value of $\Pi^1_{N/2;x}$, this leads to an even-odd modulation in both the mean and variance of  $\Pi^1_{N/2; x}$ in an ensemble of random states.
Fig.~\ref{Fig:RandEO} shows the even-odd effect on the mean of $\langle  \Pi^1_{N/2, x}\rangle$ in an ensemble of 500 parity-even random vectors in the Hilbert space, and in the central third of the eigenstates in the spectrum.  As expected, the fraction of states with $X_{N/2}=1$ (and hence $\langle  \Pi^1_{N/2, x}\rangle$) is larger for odd-length chains.
A similar modulation is observed in the standard deviation in addition to the overall trend of exponential decay with $N$.

\section{Random vectors and thermalization in the \goldchain} \label{RandomApp}

As discussed in Sec.~\ref{Sec:ETHIntro}, the ETH ansatz in Eq.~(\ref{Eq:ETHansatz}) asserts that the expectation values of local observables in an ensemble of infinite temperature eigenstates is identical the expectation values in an ensemble of random vectors in Hilbert space in the thermodynamic limit.
Here we discuss the properties of random vectors in the \goldchain Hilbert space, and show explicitly that ensembles thereof satisfy Eq.~(\ref{Eq:ETHansatz}).

We define a random vector to be a vector in the fusion tree basis $\ket{n}$ with random coefficients $\alpha_n$ that are independently and identically distributed: 
\be \label{PsiRand1}
|R \rangle = \sum_n \alpha_n |n \rangle
\ee
Because the Hamiltonian is real, we choose $\alpha_n$ to be real.
We further assume that each $\alpha_n$ is drawn from a distribution with mean zero and variance $\sigma^2$.
The variance is fixed by the requirement that the ensemble averaged norm of $\ket{R}$ be one:
\begin{align}
\overline{\bra{R} R \rangle} &= 1 \quad \Rightarrow \sum_n \overline{\alpha_n^2} =1 \\
\Rightarrow \sigma^2 &= \frac{1}{f_N} \label{Eq:Sigma2Random}
\end{align}
where $f_N$, the $N^{th}$ Fibonacci number, is the total number of states in the Hilbert space, and the overline denotes averaging with respect the random ensemble.

Consider the operator: 
\be
M_i = \frac{ 1 - \hat{x}_i }{2} \ , \ \ \ \hat{x}_i = \begin{cases} 1 \ \text{ if } X_i = 1 \\
-1  \text{ if } X_i = \tau 
\end{cases}
\ee
Then, the probability that $X_i = \tau$ in the vector $\ket{R}$ is:
\begin{align}
P(X_i = \tau) &= \langle R | M_i | R \rangle \\
& = \sum_n \alpha_n^2 \langle n | M_i | n \rangle \\ 
& = \sum_n \alpha_n^2 \delta_{X_i(n) = \tau}
\end{align}
where we have used the fact that $M_i$ is a diagonal operator in the $X_i$ basis.  
Taking the ensemble average and using Eq.~\eqref{Eq:Sigma2Random}:
\begin{align}
\overline{\langle R | M_i | R \rangle} &= \frac{1}{f_N} \sum_n  \delta_{X_i(n) = \tau} 
\end{align}
From Eq.~(\ref{PXtau}), the total number of states in the Hilbert space of an $N$-bond chain with $X_i = \tau$ is given by $f_i f_{N-i +1}$.  Thus:
\be
\overline{P(X_i = \tau)}=\overline{\langle R | M_i | R \rangle} =  \frac{f_i f_{N-i +1} }{f_N} 
\ee 
For $N \gg 1/(2\log\gmean)$, this probability is approximately:
\be 
\overline{P(X_i = \tau)} \approx \frac{\gmean}{\sqrt{5}}
\ee
at the center of the chain.
The variance of $P(X_i = \tau)$ in the random ensemble can similarly be calculated. 
For  $N \gg 1/(2\log\gmean)$, it is easily shown to be proportional to the inverse of the Hilbert space dimension:
\begin{align}
\label{Eq:VarProbXiRandom}
\overline{(\Delta P(X_i = \tau))^2} \sim \frac{1}{f_N} 
\end{align}
exactly as conjectured for a local observable by ETH (see Eq.~(\ref{Eq:ETHansatz})).

We can apply similar arguments to predict the mean and variance of the expectation values of other operators in ensembles of random vectors.  Take for example $\Pi^1_{i;2}$, the projector onto states with $Y_i =1$.  
In the fusion tree basis, we obtain:
\be \label{Yexp}
\langle R | \Pi^1_{i;2} | R\rangle =\sum_{m, n} \langle R | m^{(i)}_{abc} \rangle \langle m^{(i)}_{abc} |\Pi^1_{i;2} |n^{(i)}_{ab'c} \rangle \langle n^{(i)}_{ab'c} |R\rangle
\ee
where 
\be
|m^{(i)}_{abc} \rangle = | \ldots ,X_{i-1} = a, X_i = b, X_{i+1} = c, \ldots  \rangle
\ee
Note that $\Pi^1_{i;2}$ can only change the bond $X_i$ in the fusion tree basis (see Eq.~\eqref{YProj}).

From Eq.~(\ref{PsiRand1}):
\be \label{YrandRes}
\langle R | m^{(i)}_{abc} \rangle \langle n^{(i)}_{ab'c} | R \rangle =  \alpha_{m_{abc} } \alpha_{n_{ab'c} },
\ee
while the matrix element can be obtained from Eq.~\eqref{YProj}.
In total, this gives:
\begin{align}
\langle R | \Pi^1_{i;2} | R \rangle =\sum_{m, n}  \alpha_{m_{abc} } \alpha_{n_{ab'c} } \left \{ \delta_{b b'}  \left( \delta_{(abc) =(1 \tau 1)} \right. \right . \n
\left. + \varphi^{-2} \delta_{(abc) =( \tau 1 \tau)} + \varphi^{-1} \delta_{(abc) =(\tau \tau \tau)} \right) \n
\left. + \varphi^{- 3/2} \left(  \delta_{b' = \tau} \delta_{(abc) =( \tau 1 \tau)}   + \delta_{b'=1}  \delta_{(abc) =( \tau \tau \tau)}  \right ) \right \}  
\end{align}
On ensemble averaging, each term in the last line produces zero as $\overline{\alpha_{m_{abc} } \alpha_{n_{ab'c} }}=0$ when $b\neq b'$.
We therefore obtain:
\be
\overline{\langle R | \Pi^1_{i;2} | R \rangle } = \frac{1}{f_N}  \left( n^{(i)}_{(1 \tau 1)}+ \varphi^{-2} n^{(i)}_{( \tau 1 \tau)} +\varphi^{-1} n^{(i)}_{(\tau \tau \tau)} \right) \
\ee
where $n^{(i)}_{abc}$ is the number of states in the Hilbert space with $X_{i-1} = a, X_i =b, X_{i+1}=c$.  
This is precisely the expression for the canonical ensemble in Eq.~\eqref{Yth}. At large $N$:
\be
\overline{\langle R | \Pi^1_{i;2} | R \rangle } {\approx} \frac{1}{\varphi^2}
\ee
After an analogous calculation, we find that the variance of $\langle R | \Pi^1_{i;2} | R \rangle$ scales as the inverse of the Hilbert space dimension, similarly to Eq.~\eqref{Eq:VarProbXiRandom}.

\bibliography{shorttitles,Bibliography,biblio}

\end{document}